\documentclass{article}
\usepackage[utf8]{inputenc}

\usepackage[english]{babel}

\usepackage[sorting=none]{biblatex} 
\usepackage[letterpaper,top=2cm,bottom=2cm,left=3cm,right=3cm,marginparwidth=1.75cm]{geometry}
\usepackage{amsmath}
\usepackage{graphicx}
\usepackage[colorlinks=true, allcolors=blue]{hyperref}
\usepackage{comment}
\usepackage{csquotes}
\usepackage{eufrak}
\usepackage{adjustbox}
\usepackage{amssymb}
\usepackage{braket}
\usepackage{tikz}
\usepackage{amsfonts}
\usepackage{mathtools}
\usepackage{float}
\usepackage[affil-it]{authblk}

\title{Spin-2 Universal Minimal Solutions on Type IIA and IIB Supergravity  }
\author{Mariana Lima\footnote{m.limalp@gmail.com}}
\affil{\small International Institute of Physics, Federal University of Rio Grande do Norte,\\ 
       Campus Universit\'ario - Lagoa Nova, Natal, RN 59078-970, Brazil
}
\affil{\small Department of Theoretical and Experimental Physics, Universidade Federal do Rio Grande do Norte,\\ 
	Campus Universit\'ario - Lagoa Nova, Natal, RN 59078-970, Brazil}

\date{}

\DeclareUnicodeCharacter{0301}{\hspace{-1ex}\'{ }}
\begin{document}

\maketitle
\begin{abstract}

     In this paper, we consider the spin-2 field perturbations of four families of supergravity solutions. These include AdS$_5$ and AdS$_7$ backgrounds of type IIA as well as AdS$_4$ and AdS$_6$ backgrounds of Type IIB. As the main result, we show that, in all the cases, there is a solution given by a combination of the warp factors. We also find the respective mass spectra. 
     We analyze the normalizability of the solutions and identify the superconformal multiplets dual to them.

\end{abstract}

\section{Introduction}

The AdS/CFT conjecture, originally formulated in \cite{Maldacena:1997re}, states a correspondence between a supergravity solution in $D=10$ (or $D=11$) of the form AdS$_{p+2}\times M_{D-p-2}$, with $M$ being a compact Einstein manifold, and a superconformal field theory (SCFT) on the boundary of the AdS, that is, an SCFT in $d=p+1$.
According to the conjecture, the spectrum of the linearized fluctuations of the supergravity solution is equivalent to the spectrum of the dual SCFT in the regime of strong coupling~\cite{Gubser:1998bc,Witten:1998qj}. The conjecture thus gives non-trivial access to the structure of the SCFTs via analysis of classical supergravity equations. In this paper we will work with four families of supergravity solutions, focusing on the spin-2 perturbations along the AdS directions, which contain the modes dual to the energy-momentum tensor. We will propose a class of solutions to the linearized equations, evaluate the spectrum of dual operators and show that the found solutions are BPS.

Some relevant recent examples include the results on AdS$_2$ \cite{Lozano:2020txg}, where the authors found a new family of type IIB solutions, \cite{Lozano:2020sae}, where the authors studied a type IIA supergravity on AdS$_2\times S^3\times CY_2\times \mathcal{I}_\rho$ and proposed a dual superconformal quantum mechanics (SCQM), and \cite{Rigatos:2022ktp} where the spin-2 fluctuation of an infinity family of type IIB solutions is studied. Considerable progress was made on AdS$_3$ in the past few years, for example \cite{Eberhardt:2017fsi, Gaberdiel:2022oeu,Eberhardt:2019niq, Zacarias:2021pfz, Lima:2022hji}. On AdS$_4$ we can mention \cite{Passias:2017yke} for a study of type IIB $\mathcal{N}=2$ supergravity solutions, in \cite{ Ricardo} the authors study three type IIB supergravity solutions, where two of them are Abelian and non-Abelian T-duals of a type IIA background, and the third one is a new solution, in \cite{Akhond:2021ffz} they study an infinite family of $\mathcal{N}=4$ type IIB backgrounds on AdS$_4$ showing the holographic dual. Many works were done on AdS$_5$, like \cite{Apruzzi:2015zna}, where massive type IIA solutions on AdS$_5\times M_5$ were classified and new solutions were found; in \cite{Itsios:2019yzp} a study of spin-2 perturbation on type IIA supergravity solution was performed, considering Abelian and non-Abelian T-duals versions of AdS$_5\times S^5$, and they also studied the spectra numerically; in \cite{Gaiotto:2009gz} a study of gauge/gravity duality is done for $\mathcal{N}=2$ theories where the gravity side has the AdS$_5$ factor;
in \cite{Nunez:2019gbg} the authors study the holographic description of $\mathcal{N}=1$ and $\mathcal{N}=2$ SCFT of a supergravity solutions on AdS$_5$.
In \cite{Passias:2018swc} the authors investigated the Kaluza-Klein spectrum of a type IIA supergravity on AdS$_6$ and obtained the spin-2 modes in terms of hypergeometric functions, and in \cite{Apruzzi:2021nle} the authors studied non-supersymmetric type IIB string theory solutions; in \cite{Cremonesi:2015bld} they dwell on an infinite class of solutions on AdS$_7\times S^2\times R$ and a study of Weyl anomalies in field and gauge theories was made, while in \cite{Nunez:2018ags} the authors study the dynamics of $d=6$ superconformal field theories with $\mathcal{N}=(1,0)$, and in \cite{Bergman:2020bvi} a study of the AdS$_7$/CFT$_6$ duality was performed. In \cite{Apruzzi:2019ecr}, one can find a study of perturbative and non-perturbative stability of non-supersymmetric
solutions of type IIA with the AdS$_7$ factor.

Here we will work with some families of types IIB and IIA supergravity solutions. We are not going to discuss the details of such solutions, but focus on the study of the spin-2 field perturbation for all of them. Such a kind of perturbation is particularly interesting since the spin-2 fluctuations are known to be ``universal", they decouple from the other fields, as shown by \cite{Bachas:2011xa}, which greatly simplifies the analysis.
The first thing to notice is that we are not working with specific supergravity solutions but with infinite families of them. Each warp factor in the metrics depends on a function $V$ (which is different for each background). The function $V$, in turn, obeys a partial (or ordinary) differential equation. We will not specify any solution for $V$ before section \ref{norm}, but only its equation. Therefore, our results are valid for all possible solutions for $V$.

 In the first step, we obtain the spin-2 equations for each background. We write all these equations in the same ``universal form", namely
\begin{align}\label{UF1}
    \partial_a(p\,\partial^a\psi)+q\,\psi=-m^2\,w\,\psi,
\end{align}
where $p$, $q$, and $w$ will be written in terms of the warp factors of the metric, and $\psi$ is a part of the perturbation (we clarify this in the next section). For the solutions on AdS$_{4,5,6}$, \eqref{UF1} is an elliptic equation, and for the solutions on AdS$_7$, \eqref{UF1} is an ODE written in the standard form of a Sturm-Liouville eigenvalue problem.

The next step is to show that all the four equations we wrote as \eqref{UF1} admit simple solutions, which we refer to as ``universal minimal solutions" (UMS). Not any equation written as \eqref{UF1} has this particular solution. However, our equations do. Such solutions are expressed in terms of the warp factors. Even without using any explicit form for $V$, we found such solutions together with the corresponding spectra.
We used the word ``universal" because all four solutions can be written from the same formula in terms of their warp factors. We use the word ``minimal" because those solutions correspond to the solutions associated with the minimal eigenvalues in the papers \cite{Chen:2019ydk,Gutperle:2018wuk,Passias:2016fkm}. 
We found a spectrum for the family of solutions on AdS$_4$, which is to our knowledge not present in the existent literature.

The existence of UMS is closely connected with the supersymmetry of the supergravity solutions, and all the fluctuations organize themselves in SUSY (or superconformal) multiplets. For example, the stress-energy tensor $T_{\mu\nu}$ is annihilated by all $Q$-supercharges and is expected to be the top component of a short multiple. Determining the spectrum, we can evaluate the operator dimensions and use the results of \cite{Cordova:2016emh} to identify the specific multiplets. In \cite{Cordova:2016emh}, the authors wrote the shortening conditions (SC) for superconformal algebras in $d=3,4,5,6$ and classified superconformal multiplets. In appendix \ref{algebra} we compare our results with the results of~\cite{Cordova:2016emh} and identify the shortening conditions that our operator dimensions obey and, in particular, the supermultiplets of the stress-energy operator.

The contents of this paper are organized as follows:
In section \ref{spin2eq}, we establish our notation and, after performing a perturbation along the AdS direction, we find the spin-2 field equations for the four families of backgrounds.
In section \ref{UF and UMS}, we cast all the equations found in section \ref{spin2eq} in the same form. Then we find solutions and the respective spectrum for the fluctuations. We show that all those solutions are given by the ``same formula'', which is a combination of the warp factors. The main goal of section \ref{norm} is to check whether the UMS are normalizable. We find that all the solutions with the corresponding operator dimension fitting the classification in \cite{Cordova:2016emh} are normalizable. We investigate the conditions necessary to be imposed to have normalizability, analyticity, and orthogonality of the solutions.
In the short section \ref{3d-laplacian}, we summarized the results as a weighted Laplacian in a 3-dimensional metric. In the conclusion, section \ref{conclusion}, we summarize our results and discuss them. The Appendix \ref{f functions} translates our metric notation to the notation of the authors we are mentioning here. Finally, in Appendix \ref{algebra}, we show which of our solutions can be associated with unitary operators by finding the shortening condition (SC) obeyed by them. We also show in which multiplet the stress tensors associated with our solutions must be.

After the publication of this work on Arxiv, we became aware of the paper \cite{Rigatos:2022ktp}, where a study of an infinity family of type IIB solutions with the AdS$_2$ factor and $\mathcal{N}=4$ supersymmetry is done. Then, we extended our analysis to such families by making use of the results which are in their section 3. Therefore, in a certain sense, the AdS$_2$ results are attached solutions.

\section{Spin-2 equations}\label{spin2eq}

\subsection{Notation and procedure}\label{notation}

For all the backgrounds considered in this paper, we are going to perform a metric perturbation along the AdS direction, and we will impose that such a perturbation is transverse and traceless. Before giving more details let us fix our notation. We are going to work with 10-dimensional supergravity solutions in backgrounds as
\begin{align}\label{ads-M}
    ds^2&=G_{MN}dX^MdX^N\nonumber\\
    &=e^{2A}ds^2(AdS_n)+ds^2(\mathcal{M}_{10-n})=
    e^{2A}\Bar{g}_{\mu\nu}dx_\mu dx_\nu +
    \hat{g}_{ab}dy^ady^b,
\end{align}
where $G$ is the 10-dimensional metric, $x$ and $y$ are the AdS and the internal space coordinates, respectively, the indices $M$ and $N$ run from 1 to 10, $\mu$ and $\nu$ from 1 to $n$ (the AdS dimension), and $a$ and $b$ from 1 to $10-n$. Namely, we will work with the following backgrounds
\begin{align}
    ds^2&=f_{\text{AdS}_2}\left[
    ds^2(AdS_2)+f_Sds^2(S^2)+f_{CY}ds^2(CY_2)+f_\varphi d\varphi^2+
    f_{\mathcal{I}}d\eta^2
    \right],\label{a2}\\
    ds^2&=f_{\text{AdS}_4}\left[ds^2(\text{AdS}_4)+f_{S_1}ds^2(S_1^2)
    +f_{S_2}ds^2(S_2^2)+f_{\mathcal{I}}(d\sigma^2+d\eta^2)
    \right],\label{a4}\\
    ds^2&=f_{\text{AdS}_5}\left[ds^2(\text{AdS}_5)+f_Sds^2(S^2)+
    f_\beta d\beta^2+f_\mathcal{I}(d\sigma^2+d\eta^2)
    \right],\label{a5}\\
    ds^2&=f_{\text{AdS}_6}\left[ds^2(\text{AdS}_6)+f_Sds^2(S^2)+
    f_\mathcal{I}(d\sigma^2+d\eta^2)
    \right],\label{a6}\\
    ds^2&=f_{\text{AdS}_7}\left[ds^2(\text{AdS}_7)+f_Sds^2(S^2)+
    f_\mathcal{I}d\eta^2
    \right].\label{a7}
\end{align}
The functions $f_\text{AdS}$, $f_S$, and $f_\mathcal{I}$ are different in each line. The Calabi-Yau metric here is $\mathbb{T}^4$. Ramond and Neveu–Schwarz fields complement all these backgrounds. To see the translation between our notation and other author's notation and the expressions for the other fields, see appendix \ref{f functions}.\\

Writing the metrics as in \eqref{ads-M} is helpful to obtain the equation for $h_{\mu\nu}$, the metric perturbation. However, since we will work with four supergravity solutions, it is helpful to adopt a notation that makes it easier for us to compare our warp factors with the warps in other references.
We are going to use the following convention:
\begin{align}\label{conv}
    ds^2=f_{\text{AdS}}\left[ds^2(\text{AdS}_n)+\sum_i f_{\mathcal{M}_i}ds^2(\mathcal{M}_i)+
    f_{\mathcal{I}}ds^2(\mathcal{I})
    \right]
\end{align}
that is,
\begin{align}
&\text{AdS}_2: \mathcal{M}_1=S^2, \quad f_{\mathcal{M}_1}=f_{S},\qquad
    \mathcal{M}_2=CY_2=\mathbb{T}^4,\quad
    f_{\mathcal{M}_2}=f_{CY_2},\nonumber\\
    &\qquad \quad\mathcal{M}_3=S^1, \quad f_{\mathcal{M}_3}=f_{\varphi},\quad
    ds^2(\mathcal{I})=d\eta^2,\\
    &\text{AdS}_4: \mathcal{M}_1=S_1^2, \quad f_{\mathcal{M}_1}=f_{S_1},\qquad
    \mathcal{M}_2=S_2^2,\quad
    f_{\mathcal{M}_2}=f_{S_2},\qquad
    ds^2(\mathcal{I})=d\sigma^2+d\eta^2,\\
    &\text{AdS}_5: \mathcal{M}_1=S^2, \quad f_{\mathcal{M}_1}=f_{S},\qquad
    ds^2(\mathcal{M}_2)=d\beta^2,\quad 
    f_{\mathcal{M}_2}=f_\beta,\qquad
    ds^2(\mathcal{I})=d\sigma^2+d\eta^2,\\
    &\text{AdS}_6: \mathcal{M}_1=S^2, \quad f_{\mathcal{M}_1}=f_{S},\qquad
    ds^2(\mathcal{I})=d\sigma^2+d\eta^2,\\
    &\text{AdS}_7: \mathcal{M}_1=S^2, \quad f_{\mathcal{M}_1}=f_{S},\qquad
    ds^2(\mathcal{I})=d\eta^2.
\end{align}

Usually, the warp factors are labelled by $f_1$, $f_2$, etc. (see appendix \ref{f functions}). However, what each author calls $f_1$ is different from the others, so we are going to use the notation above to make it easier to compare what is written here with their papers and to be able to summarize our results. In this notation we are splitting the solutions into three parts: AdS, $\mathcal{I}$, and the rest ($\mathcal{M}_i$). The warps do not depend on all coordinates, but only on the coordinates of $\mathcal{I}$, and that is how we define which part of the metric is $\mathcal{I}$.\\

In this project, we will always work in the Einstein frame. We will alternate between the notations $\Psi$ and $\Psi_n$, where $\Psi_n$ is the $\Psi$ function for the AdS$_n$ case. The same applies to $\psi$ and $\psi_n$, the function which appears after to expand $\Psi$ in spherical harmonics.\\

By performing a perturbation along the AdS direction as
\begin{align}
    \Bar{g}_{\mu\nu}\rightarrow \Bar{g}_{\mu\nu}+\delta \Bar{g}_{\mu\nu}, \qquad
    \delta g_{\mu\nu}=e^{2A}h_{\mu\nu}=f_{\text{AdS}}h_{\mu\nu},
    \qquad \text{and}\qquad \delta \hat{g}_{ab}=0,
\end{align}
one finds that $h_{\mu\nu}$ obeys the Laplace-Beltrami equation\footnote{The detailed calculations are done in \cite{Lima:2022hji}}
\begin{align}
    \frac{1}{\sqrt{G}}\partial_M(\sqrt{G}G^{MN}\partial_N)h_{\mu\nu}=0.
\end{align}
After separating the AdS and $\mathcal{M}_{10-n}$ coordinates,
$h_{\mu\nu}=h_{\mu\nu}^{tt}\Psi(y)$, and imposing the transverse-traceless condition, the linearized Einstein equations give \cite{Bachas:2011xa, Lima:2022hji}
\begin{align}\label{s2}
\left[
    \frac{e^{(2-n)A}}{\sqrt{\hat{g}}}\partial_{y_a}
    \left(e^{nA}\sqrt{\hat{g}}\hat{g}^{ab}\partial_{y_b}\right)
\right]\Psi=-m^2\Psi.
\end{align}

\subsection{The supergravity solutions and the respective spin-2 equations}\label{s2-equations}

Here we are going to work with two families of type IIB supergravity solutions, one on AdS$_4$ \cite{Ricardo} and one on AdS$_6$ \cite{Legramandi:2021uds}, and two of type IIA, one on AdS$_5$ \cite{Itsios:2019yzp}, and one on AdS$_7$ \cite{Cremonesi:2015bld, Nunez:2018ags}\footnote{Although we did the calculations on both conventions, in the present paper we are going to use the same convention as in \cite{Nunez:2018ags} for the AdS$_7$ case.}.

The warp factors in \eqref{a4} to \eqref{a7} are written in terms of potential functions, $V$, which depend on the coordinates of $\mathcal{I}$. Each solution has its potential function $V$, and each $V$ obeys a PDE (ODE for the solution on AdS$_7$).
To see the warp factors in terms of $V$, as well as the dilaton and other fields, see \cite{Ricardo, Legramandi:2021uds, Itsios:2019yzp, Nunez:2018ags}. The equations for the $V$ functions will be important:
\begin{align}
    &\frac{d^2V_2}{d\eta^2}=0,\\
    &\partial_\sigma\left[\sigma^2\partial_\sigma V_4(\sigma,\eta)\right]+\sigma^2\partial_\eta^2V_4(\sigma,\eta)=0,\label{V4}\\
    &\sigma\partial_\sigma(\sigma\partial_\sigma
    V_5(\sigma,\eta)+
    \sigma^2\partial_\eta^2V_5(\sigma,\eta)=0,\label{V5}\\
    &\partial_\sigma\left[\sigma^2\partial_\sigma V_6(\sigma,\eta)\right]+\sigma^2\partial_\eta^2V_6(\sigma,\eta)=0,\label{V6}\\
    &\partial_\eta^3V_7(\eta)=-162\pi^3F_0\label{V7},
\end{align}
where $F_0$ is a piecewise constant function.\\

We are not going to write every single step for each background, however, two of them are the same for all solutions we cover here, namely: 

1) expanding $\Psi$ in spherical harmonics;

2) multiplying the result equation by the function $w$ (we will soon show which function is it).\\

By 1) we mean the following:

\begin{align}\label{Psi_exp}
\begin{cases}
\vspace{0.1in}
&\Psi_2=\sum_{l,m,n,p}Y_l^m(\theta,\phi)e^{in . \theta}e^{ip\varphi}
\psi_{l,m,n,p}(\eta),\quad
n=\{n_1,n_2,n_3,n_4\},\quad n_i\in\mathbb{Z},\\
& \Psi_4=
    \sum_{l_1,m_1,l_2,m_2}Y_{l_1}^{m_1}(\theta_1,\phi_1)
    Y_{l_2}^{m_2}(\theta_2,\phi_2)\psi_{m_1,m_2}^{l_1,l_2}(\sigma,\eta),\\
    \vspace{0.1in}
&\Psi_5=\sum_{l,m,n}Y_l^m(\theta,\phi)e^{in\beta}\psi_{l,m,n}(\sigma,\eta),
    \qquad n\in \mathbb{Z},\\
    \vspace{0.1in}
    & \Psi_{6}=\sum_{l,m}Y_l^m(\theta,\phi)\psi_{l,m}(\sigma,\eta),\\
    & \Psi_{7}=\sum_{l,m}Y_l^m(\theta,\phi)\psi_{l,m}(\eta).
\end{cases}
\end{align}
Note that $\Psi_5$ has an exponential term with the variable $\beta$. The expansions \eqref{Psi_exp} are merely the separation of variables procedure.

After 1), each $\psi$ above obeys an equation with the following format
\begin{align}\label{g}
     g_2(\sigma,\eta)\partial_\sigma^2\psi+g_2(\sigma,\eta)
    \partial_\eta^2\psi+
    g_{10}(\sigma,\eta)\partial_\sigma\psi+
    g_{01}(\sigma,\eta)\partial_\eta\psi+
    \left[g_0(\sigma,\eta)+m^2\right]\psi = 0.
\end{align}

Then, the $w$ function mentioned in 2) is
\begin{equation}\label{ww}
    w=\frac{1}{g_2}\exp{\left[ 
    \int \frac{g_{10}}{g_2}d\sigma\right]}.
\end{equation}

After 2), equation \eqref{g} becomes
\begin{align}
    \partial_a(p(\sigma,\eta)\partial^a\psi)+q(\sigma,\eta)\psi=
    -m^2w(\sigma,\eta)\psi,\qquad
    a=\sigma, \eta.
\end{align}

\subsubsection{Type IIB}\label{typeIIB}

\paragraph{AdS$_2$}

Since the $\mathcal{I}$-geometry in \eqref{a2} is one-dimensional, after applying \eqref{s2} the resulting equation for $\psi_2$ is an ODE in the Sturm-Liouville standard form:
\begin{align}\label{psi2}
    \frac{d}{d\eta}\left(u^2\frac{d\psi}{d\eta}
    \right)+\left[-4l(l+1)\hat{h}_4h_8-u(n^2h_8+p^2u)-l(l+1)u'^2
    \right]\psi=-4m^2\hat{h}_4h_8,
\end{align}
where $u$, $\hat{h}_4$, and $h_8$ are functions of $\eta$, and $n^2=\sum_{i=1}^4n_i^2$. For detailed calculations see \cite{Rigatos:2022ktp}. In particular, $u(\eta)=V_2(\eta)$.

\paragraph{AdS$_4$}

Applying \eqref{s2} for the background \eqref{a4}, one finds

\begin{align}\label{Psi4pde}
    &\left[ \frac{f_{\text{AdS}}}{f_{S_1}}\nabla^2_{(S_1^2)}+
    \frac{f_{\text{AdS}}}{f_{S_2}}\nabla^2_{(S_2^2)}\right]\Psi_4+
    \frac{1}{f_\mathcal{I}}\left[\sum_{i=\sigma,\eta}\left( 2\partial_i f_\text{AdS}+
    f_\text{AdS}\frac{\partial_i f_{S_1}}{f_{S_1}}+
    f_\text{AdS}\frac{\partial_i f_{S_2}}{f_{S_2}}\right)\partial_i
    \right]\Psi_4\\
    &+\frac{f_\text{AdS}}{f_\mathcal{I}}\left(
    \partial_\sigma^2+\partial_\eta^2
    \right)\Psi_4=-m^2\Psi_4.\nonumber
\end{align}
The solutions for the angular parts are spherical harmonics.
After writing $\Psi$ as
\begin{align}\label{Psi4}
    \Psi_4(\theta_1,\phi_1,\theta_2,\phi_2,\sigma,\eta)=
    \sum_{l_1,m_1,l_2,m_2}Y_{l_1}^{m_1}(\theta_1,\phi_1)
    Y_{l_2}^{m_2}(\theta_2,\phi_2)\psi_{m_1,m_2}^{l_1,l_2}(\sigma,\eta),
\end{align}
one finds
\begin{align}\label{psi4-1a}
&\partial_\sigma^2\psi+\partial_\eta^2\psi+
\left(4\frac{\partial_\sigma f_\text{AdS}}{f_\text{AdS}}+
\frac{\partial_\sigma f_{S_1}}{f_{S_1}}+
\frac{\partial_\sigma f_{S_2}}{f_{S_2}}
\right)\partial_\sigma \psi+
\left(4\frac{\partial_\eta f_\text{AdS}}{f_\text{AdS}}+
\frac{\partial_\eta f_{S_1}}{f_{S_1}}+
\frac{\partial_\eta f_{S_2}}{f_{S_2}}
\right)\partial_\eta \psi+\nonumber\\
&    \left(m^2+\frac{\kappa_{S_1}}{f_{S_1}}+
    \frac{\kappa_{S_2}}{f_{S_2}}
    \right)f_\mathcal{I}\psi=0.
\end{align}
In other words, we have equation \eqref{g} with
\begin{align}
    &g_2(\sigma,\eta)=f_\mathcal{I}^{-1},\\
    &g_{10}(\sigma,\eta)=\left(4\frac{\partial_\sigma f_\text{AdS}}{f_\text{AdS}}+
\frac{\partial_\sigma f_{S_1}}{f_{S_1}}+
\frac{\partial_\sigma f_{S_2}}{f_{S_2}}
\right)f_\mathcal{I}^{-1},\\
&g_{01}(\sigma,\eta)=\left(4\frac{\partial_\eta f_\text{AdS}}{f_\text{AdS}}+
\frac{\partial_\eta f_{S_1}}{f_{S_1}}+
\frac{\partial_\eta f_{S_2}}{f_{S_2}}
\right)f_\mathcal{I}^{-1},\\
 &g_0(\sigma,\eta)=\left(\frac{\kappa_{S_1}}{f_{S_1}}+
    \frac{\kappa_{S_2}}{f_{S_2}}
    \right)f_\mathcal{I}^{-1},
\end{align}
and the $w$ function defined in step 2 is
\begin{align}
    w(\sigma,\eta)=f_\text{AdS}^4f_\mathcal{I}f_{S_1}f_{S_2}.
\end{align}

After applying 2) in the equation \eqref{psi4-1a}, one finds that $\psi_{m_1,m_2}^{l_1,l_2}$ obeys
\begin{align}\label{psi4}
    \partial_a (f_\text{AdS}^4f_{S_1}f_{S_2}\, \partial^a\psi_{m_1,m_2}^{l_1,l_2})+
    f_\text{AdS}^4f_{\mathcal{I}}(\kappa_{S_1}f_{S_2}+
    \kappa_{S_2}f_{S_1})\,\psi_{m_1,m_2}^{l_1,l_2}=
    -m^2f_\text{AdS}^4f_{S_1}f_{S_2}f_{\mathcal{I}}\,\psi_{m_1,m_2}^{l_1,l_2},
\end{align}
where $\kappa_{S_i}=-l_i(l_i+1)$, $a=\sigma,\eta$. \\

\paragraph{AdS$_6$}

When we perturb \eqref{a6} along AdS$_6$ following the procedure described at the beginning of this section, the resulting equation for $\Psi_6$ is given by
\begin{equation}
\left[
    \frac{1}{f_S}\nabla^2_{S^2}+\frac{1}{f_\mathcal{I}}\left(
    4\frac{\partial_\eta f_\text{AdS}}{f_\text{AdS}}\partial_\eta +
    \frac{\partial_\sigma f_\text{AdS}}{f_\text{AdS}}\partial_\sigma+
    \frac{\partial_\eta f_S}{f_S}\partial_\eta +
    \frac{\partial_\sigma f_S}{f_S}\partial_\sigma
    \right)\right]\Psi_6
    =-m^2\Psi_6.
\end{equation}

The solution in AdS$_6$ \eqref{a6} has only one sphere, and then $\Psi_6$ is expanded as showed in \eqref{Psi_exp}.
After steps 1) and 2), we obtain
\begin{align}\label{psi6}
    \partial_a (f_\text{AdS}^4f_{S}\, \partial^a\psi_{l,m})+
    \kappa_S f_\text{AdS}^4f_{\mathcal{I}}\,\psi_{l,m}=
    -m^2f_\text{AdS}^4f_{S}f_{\mathcal{I}}\,\psi_{l,m},\qquad
    \kappa_S=-l(l+1).
\end{align}

\subsubsection{Type IIA}

\paragraph{AdS$_5$}

We start the calculation as described in section \ref{notation}, that is, we perform a perturbation along the AdS direction, separate variables, and impose the traceless and transverse conditions. All the details one can find in \cite{Itsios:2019yzp}. One finds the following equation for $\Psi_5$

\begin{align}
    \left[\frac{1}{f_S}\nabla^2_{S^2}+
    \frac{1}{f_\beta}\partial_\beta^2+
    \frac{1}{f_\mathcal{I}}\sum_{i=\sigma,\eta}\left(
    4\frac{\partial_if_\text{AdS}}{f_\text{AdS}}\partial_i+
    \frac{\partial_if_S}{f_S}\partial_i+\frac{1}{2}
    \frac{\partial_if_\beta}{f_\beta}+ \partial^2_i\right)
    \right]\Psi_5=-m^2\Psi_5.
\end{align}
This time the function $\Psi$ is expanded as
\begin{align}
    \Psi_5=\sum_{l,m,n}Y_{l}^{m}(\theta,\phi)e^{in\beta}
    \psi_{l,m,n}(\sigma,\eta),\qquad n\in\mathbb{Z}.
\end{align}
And then we find the following PDE for $\psi_{l,m,n}$
\begin{align}
     \partial_a (f_\text{AdS}^4f_S\sqrt{f_\beta}\, \partial^a\psi_{l,m,n})+
     \frac{f_\text{AdS}^4f_\mathcal{I}}{\sqrt{f_\beta}}
     (\kappa_S f_\beta+\kappa_\beta f_S)
     \,\psi_{l,m,n}=-m^2f_\text{AdS}^4f_S\sqrt{f_\beta}f_\mathcal{I}\,\psi_{l,m,n}\label{psi5}
\end{align}
with $\kappa_S=-l(l+1)$, $\kappa_\beta=-n^2$.\\

\paragraph{AdS$_7$}

The background \eqref{a7} is a particular case since the warp factors depend on only one variable, $\eta$. The resulting equation for $\Psi_7$ is

\begin{align}
\left[
    \frac{1}{f_S}\nabla^2_{(S^2)}+\frac{1}{f_\mathcal{I}}\frac{d^2}{d\eta^2}+
    \frac{1}{f_\mathcal{I}}\left(
    4\frac{f_\text{AdS}'}{f_\text{AdS}}+\frac{ f_S '}{f_S}
    -\frac{1}{2}\frac{ f_\mathcal{I}'}{f_\mathcal{I}}
    \right)\frac{d}{d\eta}
    \right]\Psi_7=-m^2\Psi_7.
\end{align}
Since \eqref{a7} has only one sphere, we expand $\Psi_7$ as
\begin{align}
    \Psi_7 = \sum_{l,m}Y_l^m(\theta,\phi)\psi_{l,m}(\eta),
\end{align}
and then we obtain
\begin{align}\label{psi7}
    \frac{d}{d\eta}\left(
    \frac{f_\text{AdS}^4f_S}{\sqrt{f_\mathcal{I}}}\frac{d\psi_{l,m}}{d\eta}
    \right)+\kappa_S f_\text{AdS}^4\sqrt{f_I}\,\psi_{l,m}=
    -m^2f_\text{AdS}^4f_S\sqrt{f_I}\,\psi_{l,m},\qquad \kappa_S=-l(l+1).
\end{align}
The equation above, together with the appropriate boundary conditions form a Sturm-Liouville eigenvalue problem.

\section{Universal form and the universal minimal solutions}\label{UF and UMS}
For each case we are considering, the equations can be written in the following form:
\begin{align}\label{UF}
    \partial_a (p\, \partial^a\psi)+q\,\psi=-m^2w\,\psi,
\end{align}
where $a=\sigma,\, \eta$  are coordinates of the $\mathcal{I}$ geometry, and $p$, $q$, and $w$ depend only on such variables. The equation above is an elliptic differential equation, and therefore the possible values of $m^2$ are the spectra of an elliptic operator\footnote{Note that for the AdS$_7$ case, instead of a PDE we have an ODE since the warp factors depend only on one variable. And then, instead of an elliptic equation, we have a Sturm-Liouville eigenvalue problem.} with weight function $w$. The spin-2 equations obtained in the previous sections can be summarized as \eqref{UF}, where $p$, $q$, and $w$ are given by

\begin{align}\label{all-to}
w=f_{\text{AdS}}^4f_{\mathcal{I}}^{d_{\mathcal{I}}/2}\prod_if_{\mathcal{M}_i}^{d_i/2},\qquad
    p=\frac{w}{f_{\mathcal{I}}},\qquad
    q=w\, \sum_i\frac{\kappa_i}{f_{\mathcal{M}_i}}
\end{align}
where $d_\mathcal{I}$ is the dimension of $\mathcal{I}$, $d_i$ is the dimension of $\mathcal{M}_i$, and $\kappa_i$ is the quantum number (separation constant) associated to $\mathcal{M}_i$.

The idea to obtain the universal minimal solution is quite simple: we write the function $\psi$ as $\psi=p^\alpha\chi(\sigma,\eta)$ in the hope that this substitution will simplify equation \eqref{all-to} in such a way that it helps us to find a solution. ``Simplify" is not a good word here because the result is a complicated PDE for $\chi(\sigma,\eta)$, but it gives us an idea of how to find a solution for $\psi$. The PDE for $\chi(\sigma,\eta)$ has the following form
\begin{equation}\label{chi}
h_0\chi+h_{10}\partial_\sigma\chi+h_{01}\partial_\eta\chi
    h_{20}\partial_\sigma^2\chi+h_{02}\partial_\eta^2\chi=0,
\end{equation}
where $h_0,\, h_{01},\,h_{10},\, h_{02}$, and $h_{20}$ depend on $\sigma$ and $\eta$, and, in particular, $h_{01}$ and $h_{10}$ are proportional to $2\alpha+1$. Then, the first idea is to make $\alpha=-1/2$ to reduce \eqref{chi} to a PDE with second and zeroth derivatives only. The second idea is to explore the function $h_0$, which is the only function that carries the constants $m^2$ and $\kappa_i$. If it is possible to rewrite $h_0$ as
\begin{equation}\label{iota}
    h_0(\sigma,\eta)=(m^2-\iota)h(\sigma,\eta),
\end{equation}
then $\chi(\sigma,\eta)=\chi_0=$ constant solves \eqref{chi} when $m^2=\iota$, where $\iota$ is a constant, a combination of $\alpha$ and $\kappa_i$.

We followed the second idea and we did not find that $h_0$ can be written as \eqref{iota}, but we found something similar to it in our four cases. 
To put it shortly, we found that
\begin{equation}\label{ums}
    \psi(\sigma,\eta)=p(\sigma,\eta)^\alpha\sigma^\gamma
\end{equation}
is a solution of \eqref{UF} when $m^2$ is an specific combination of the constants $\alpha$, $\gamma$, and the separation constants $\kappa_i$, and \eqref{ums} is what we are calling \textit{universal minimal solution} (UMS).

Now let us shed some light on how this calculation is done.
After substituting \eqref{ums} in \eqref{UF}, we replace the functions $p$, $q$, and $w$ by their expressions given in \eqref{all-to}, and then we need to replace the parts $f_\text{AdS}$, $f_{\mathcal{I}}$, and $f_{\mathcal{M}_i}$ by their expressions in terms of the $V$ functions given in the appendix \ref{f functions}. Also, we are not going to use any explicit solution for $V$, however, we will use their PDEs.

The discussion about the normalizability of the solutions will be done in the next section.

\subsection{Finding the UMS}

\paragraph{AdS$_4$}

We will try to clarify what we just said by doing the AdS$_4$ case. For the others, we will only write the results. The function in equation \eqref{psi4} are $f_\text{AdS}$, $f_\mathcal{I}$, $f_{S_1}$, and $f_{S_2}$. The $\kappa_i$ constants are given by $\kappa_1=-l_1(l_1+1)$ and $\kappa_2=-l_2(l_2+1)$. We substitute the function $\psi_{m1,m2}^{l_1,l_2}$ by
\begin{equation}
    \psi_4(\sigma,\eta)=p(\sigma,\eta)^\alpha=\left[
    f_\text{AdS}^4f_{S_1}f_{S_2}
    \right]^\alpha.
\end{equation}
Then, we substitute $f_\text{AdS}$, $f_\mathcal{I}$, $f_{S_1}$, and $f_{S_2}$ by their expressions given in the appendix \ref{f functions} (see the table). 
To get rid of $\partial_\sigma^2\partial_\eta V$, for example, we derive \eqref{V4} by $\eta$ obtaining
\begin{align}\label{deV4}
    \partial_\sigma^2\partial_\eta V=
    -\partial_\eta^3V-\frac{2\partial_\sigma\partial_\eta V}{\sigma}.
\end{align}
When we substitute $\partial_\sigma^2\partial_\eta V$ by the right-hand side of the equation above, we also get rid of the $\partial_\eta^3V$ which was already there. After some manipulations, we find\footnote{The question mark sign is there because we are trying to make it zero.}
\begin{align}
    (m^2-4\alpha(3+4\alpha))(\partial_\eta V)^2+
    \left[l_1+l_1^2-2\alpha(1+2\alpha)
    \right]\sigma^2\left[(\partial_\eta V)^2+(\partial_\sigma\partial_\eta V)^2
    \right]\nonumber\\+
    \left[m^2+l_1+l_1^2-l_2-l_2^2-4\alpha(3+4\alpha)
    \right]\sigma\partial_\eta V\partial_\sigma\partial_\eta V
    \stackrel{\text{?}}{=}0.
\end{align}
If the three coefficients above vanish, we have what we want. Solving the system
\begin{align}
    m^2-4\alpha(3+4\alpha)&=0,\nonumber\\
    l_1+l_1^2-2\alpha(1+2\alpha)&=0,\\
    m^2+l_1+l_1^2-l_2-l_2^2-4\alpha(3+4\alpha)&=0,\nonumber
\end{align}
we find
\begin{align}
    \l_1=l_2\equiv l,\qquad
    &\alpha=\frac{l}{2}\qquad\qquad\,\text{and}\qquad
    m^2=2(2l^2+3l),\qquad\text{or}\\
    &\alpha=-\frac{l+1}{2}\qquad\text{and}\qquad
    m^2=2(2l^2+l-1).
\end{align}
Therefore, we found two solutions and their respective spectrum.

Now we use the same procedure, but replacing  $\psi_{m1,m2}^{l_1,l_2}$ by
\begin{equation}\label{psi4gamma}
    \psi_4(\sigma,\eta)=p(\sigma,\eta)^\alpha\sigma^\gamma=\left[
    f_\text{AdS}^4f_{S_1}f_{S_2}
    \right]^\alpha\sigma^\gamma.
\end{equation}
After substituting \eqref{psi4gamma} in \eqref{psi4}, we do some simplifications and use \eqref{deV4} again, then we obtain
\begin{align}
    \left[4\alpha(3+4\alpha)-m^2\right]\left(\partial_\eta V\right)^2
        +
    \left[4\alpha(3+4\alpha+2\gamma)+2\gamma+l_2(l_2+1)-l_1(l_1+1)\right]
    \sigma\partial_\eta V\partial_\sigma\partial_\eta V\nonumber\\
    -(l_1-2\alpha-\gamma)(1+l_1+2\alpha+\gamma)\sigma^2\left[
    (\partial_\eta^2 V)^2+(\partial_\sigma\partial_\eta V)^2
    \right]     \stackrel{\text{?}}{=}0.
\end{align}
This time, we find four solutions:
\begin{align}
&\alpha= \frac{l^+}{4},\qquad\,\,\,\,\,\,\,\,\qquad\quad\gamma=-\frac{l^-}{2},\qquad\qquad
   \,\, m^2=l^+(l^++3),\label{g1}\\
   &\alpha=-\frac{l^++2}{4},\qquad \qquad \gamma=\frac{l^-}{2},\qquad\qquad\quad
   m^2=l^+(l^++1)-2,\label{g2}\\
   &\alpha=-\frac{l^-+1}{4},\qquad\qquad \gamma=\frac{l^++1}{2},\qquad\quad\,\,
   m^2=l^-(l^--1)-2,\label{g3}\\
   &\alpha = \frac{l^--1}{4},\,\,\;\,\,\qquad\qquad \gamma=-\frac{l^++1}{2},\qquad\,\,
   m^2 = l^-(l^-+1)-2,\label{g4}
\end{align}
where $l^+=l_1+l_2$ and $l^-=l_2-l_1$. When $l_1=l_2$, \eqref{g1} and \eqref{g2} are exactly the previous solutions. Since $l_1$ and $l_2$ are non negative integers, either $\gamma$ in \eqref{g3} or in \eqref{g4} can vanish. Solutions \eqref{g3} and \eqref{g4} can be summarized as
\begin{align}
    \alpha^\pm=\pm\frac{l^-\mp 1}{4},\qquad
    \gamma^\pm= \mp\frac{l^++1}{2},\qquad
    m^2_\pm = l^-(l^-\pm 1)-2.
\end{align}

We wrote four solutions, but they are all connected via $l_i\rightarrow-l_i-1$. Therefore, using \eqref{g1} as a reference, we have
\begin{align}
    \eqref{g1} \xrightarrow[l_1\rightarrow-l_1-1]{} \eqref{g4},
    \qquad\quad
    \eqref{g1} \xrightarrow[l_2\rightarrow-l_2-1]{} \eqref{g3},
    \qquad\quad
    \eqref{g1} \xrightarrow[l_1\rightarrow-l_1-1]{l_2\rightarrow-l_2-1} \eqref{g2}.
\end{align}

The mass $m^2$ connects with the operator dimension $\Delta$ via \cite{Penedones:2016voo}
\begin{align}\label{delta}
    \Delta=\frac{d-1}{2}+\sqrt{\frac{(d-1)^2}{4}+m^2}.
\end{align}
Therefore, for the four solutions above, we find
\begin{align}\label{delta-4}
    \Delta_1&=l^++3,\qquad
    \Delta_2=2+l^+,\qquad
    \Delta_3^+=\frac{1}{2}(3+|l^--1|)  \qquad
    \Delta_3^-=2-l^-,  \nonumber\\
    \Delta_4^+&=l^-+2, \qquad \Delta_4^-=1-l^-,
\end{align}
where the upper labels $\pm$ in $\Delta$ indicate the sign of $l^-$.


\paragraph{AdS$_5$}

Now let us move to the solution on AdS$_5$:
\begin{align}\label{psi5gamma}
    \psi_5(\sigma,\eta)=p(\sigma,\eta)^\alpha \sigma^\gamma=
   \left[ f_{\text{AdS}}^4f_S\sqrt{f_\beta}\right]^\alpha\sigma^\gamma.
\end{align}
We plug it in \eqref{psi5}, substitute $f_{\text{AdS}}$, $f_S$, $f_\beta$, and $f_\mathcal{I}$ by their expressions in terms of $V_5$ as given in the appendix \ref{f functions}, and use \eqref{V5} and its derivative with respect to $\sigma$. After some manipulations we find
\begin{align}
    H(\sigma,\eta)\left\{
    \left[
    2l(l+1)-n^2-m^2+2\alpha+2(3\alpha+\gamma)(2+3\alpha+\gamma)
    \right] (\partial_\sigma V)^2+\right. \nonumber\\
    \left[
    n^2-m^2+4(5\alpha+6\alpha^2+\gamma+2\alpha\gamma)
    \right]\sigma\partial_\sigma V\partial_\sigma^2 V\nonumber\\
    \left. -2(l-2\alpha)(1+l+2\alpha)\sigma^2\left[
    (\partial_\sigma^2V)^2+\sigma^2
    (\partial_\eta^3V + \partial_\sigma^2\partial_\eta V)^2
    \right] \right\}=0,
\end{align}
where $H(\sigma,\eta)$ is a combination of $V_5(\sigma,\eta)$, its derivatives and powers of $\sigma$.
Therefore, when $\alpha$, $\gamma$, and $m^2$ solve the system
\begin{align}
    2l(l+1)-n^2-m^2+2\alpha+2(3\alpha+\gamma)(2+3\alpha+\gamma)&=0\\
     n^2-m^2+4(5\alpha+6\alpha^2+\gamma+2\alpha\gamma)&=0\\
     -2(l-2\alpha)(1+l+2\alpha)&=0
\end{align}
\eqref{psi5gamma} is a solution of \eqref{psi5}. The system above is solved by
\begin{align}
    &\alpha=\frac{l}{2},\qquad \qquad\,\,\gamma_\pm=-\frac{l}{2}\pm n,\qquad\quad
    m^2_\pm=(2l\pm n)(2l\pm n+4),\qquad\text{or}\label{g51}\\
    &\alpha=-\frac{l+1}{2},\qquad \gamma_\pm=\frac{l+1}{2}\pm n,\qquad
   \, m^2_\pm=(2l\mp n-2)(2l\mp n+2),\label{g52}
\end{align}
where\footnote{We can (and we will) fix $n$ to be positive. Therefore, from now on $n\in\mathbb{Z}_{\geq 0}$.} $n\in\mathbb{Z}_{\geq 0}$. Note that for \eqref{g51}, $\gamma_\pm$ can vanish only when $l$ is even, while for \eqref{g52}, only when $l$ is odd. When $l\rightarrow-l-1$ in \eqref{g51}, the result is \eqref{g52}. Those solutions were found in \cite{Chen:2019ydk}.

The operator dimensions corresponding to each solution are then
\begin{align}
    \Delta_1^+=2l+n+4,\qquad
    \Delta_1^-=2+|n-2l-2|,\qquad
     \Delta_2^+=2+|n-2l|,\qquad
     \Delta_2^-=n+2l+2.\qquad
\end{align}

When $\gamma_\pm=0$ the solutions reduce to
\begin{align}
    &\alpha=\frac{l}{2}\quad\qquad\,\text{and}\quad n^2=\frac{l^2}{4},\quad
    \qquad\,\,\,\,
     \quad\quad m^2=\frac{5}{4}l(5l+8),\label{n-even}\\
     &\alpha=-\frac{l+1}{2}\quad\text{and}\quad n^2=\frac{(l+1)^2}{4},\qquad\quad
     m^2=\frac{5}{4}(5l^2+2l-3).\label{n-odd}
\end{align}
Therefore, we found solutions for $n$ even and odd
\begin{align}
    &n \,\,\text{even}\quad\rightarrow\quad
    \psi(\sigma,\eta)=\left(f_{\text{AdS}}^4f_S\sqrt{f_\beta}
    \right)^{l/2},\qquad\quad
    m^2=\frac{5}{4}l(5l+8)=k(k+4),\quad k=\frac{5l}{2},\\
    &n \,\,\text{odd}\,\,\quad\rightarrow\quad
    \psi(\sigma,\eta)=\left(f_{\text{AdS}}^4f_S\sqrt{f_\beta}
    \right)^{-(l+1)/2},\quad
    m^2=\frac{5}{4}(5l^2+2l-3)=k(k+4),\quad k=-\frac{5}{2}(l+1).
\end{align}
Note that for $n$ odd, the lowest value for $m^2$ is $m^2=-15/4$ (when $l=0$).\\


\paragraph{AdS$_6$} For the background \eqref{a6} we found that
\begin{align}
    \psi_6(\sigma,\eta)=p(\sigma,\eta)^\alpha\sigma^\gamma=
    (f_\text{AdS}^4f_S\sqrt{f_\beta})^\alpha\sigma^\gamma
\end{align}
is solution for
\begin{align}
    &\alpha=\frac{l}{2},\qquad \gamma=0,\qquad m_1^2=3l(3l+5),\qquad\text{or}\\
    &\alpha=\frac{l}{2},\qquad\gamma=1, \qquad m^2_2=(3l+1)(3l+6),
\end{align}
and, when we perform $l\rightarrow-l-1$,
\begin{align}
    &\alpha=-\frac{l+1}{2},\qquad\gamma=0,\qquad m^2_3=(3l-2)(3l+3)\qquad\text{or}\\
     &\alpha=-\frac{l+1}{2},\qquad\gamma=1,\qquad m^2_4=(3l-3)(3l+2),
\end{align}
and the $\Delta$'s are
\begin{align}
    \Delta_1=3l+5,\qquad \Delta_2=3l+6,\qquad
    \Delta_3=3l+3,\qquad \Delta_4 = \frac{1}{2}(5+|1-6l|),
\end{align}
which is in agreement with \cite{Gutperle:2018wuk}.
All the $m^2$'s and $\Delta$'s above are positive, but $m_3^2$ and $m_4^2$ when $l=0$. All the square masses are written in the form $\kappa(\kappa+5)$.\\

\paragraph{AdS$_7$}
As we mentioned before, the AdS$_7$ case is different because instead of a partial, we obtain an ordinary differential equation since $\mathcal{I}$ is 1-dimensional. As a consequence, to write the ODE in a compact form this time means to write it as a Sturm-Liouville eigenvalue problem, not an elliptic equation as in the previous examples. Even so, we can apply the same procedure. However, it does not make sense anymore to write a solution as $\psi(\eta)=p(\psi)^\alpha\sigma^\gamma$, since $\psi_7$ depends only on $\eta$. One could then try a solution like $\psi_7(\eta)=p(\eta)\eta^\gamma$, but we would find that this function is a solution of \eqref{psi7} only for $\gamma=0$. In the appendix \ref{f functions}, $\alpha$ is a function of $\eta$. Then, to avoid confusion with the notation, this time we will use another letter for the exponent. The function
\begin{equation}
    \psi_7(\eta)=p(\eta)^\lambda=\left(\frac{f_\text{AdS}^4f_S}{\sqrt{f_\mathcal{I}}}
    \right)^\lambda
\end{equation}
solves \eqref{psi7} when
\begin{align}
    &\lambda=\frac{l}{2}\qquad\text{and}\qquad
    m_1^2=4l(4l+6), \qquad\text{or}\label{m7}\\
    &\lambda=-\frac{l+1}{2},\qquad\text{and}\qquad
    m^2_2=(4l-2)(4l+4)
\end{align}

As the reader is probably guessing, when we write $\psi$ as $\psi(\eta)=p(\eta)^\alpha\chi(\eta)$, we find that this is a solution with $\chi(\sigma,\eta)=1$ when $\alpha$ is either $l/2$ or $-(l+1)/2$, together with  with the conditions for the mass. Summarizing, we have
\begin{align}
    &\psi(\eta)=p(\eta)^{l/2}=
    \left(\frac{f_\text{AdS}^4f_S}{\sqrt{f_\mathcal{I}}}
    \right)^{l/2},\qquad\,\, \qquad\quad\text{with}\quad
    m^2=4l(4l+6),\label{ads7-tomasiello}\\
    &\psi(\eta)=p(\eta)^{-(l+1)/2}=
    \left(\frac{f_\text{AdS}^4f_S}{\sqrt{f_\mathcal{I}}}
    \right)^{-(l+1)/2}, \quad\text{with}\quad
    m^2=(4l-2)(4l+4),
\end{align}
and the $\Delta$'s are
\begin{equation}
    \Delta_1=4l+6,\qquad\Delta_2=4l+4.
\end{equation}

The solution \eqref{ads7-tomasiello} was found in \cite{Passias:2016fkm}.

In principle, our goal is to find out the universal solutions for each case considered. However, we can point out something more about the equation for $\psi_7$.
The solution for $V_7$ is a piecewise third-order polonium since $F_0$ is a piece-wise constant function. Then, after replacing the functions $f_\text{AdS}$, $f_S$, and $f_\mathcal{I}$ by their expression in terms of $V_7$ (denoted by $\alpha$ in the Appendix \ref{f functions}), we find that \eqref{ads7-tomasiello} is simply a piecewise polynomial in $\eta$. We can go far: the ODE for $\psi_7$ is simple enough to be completely solved. Actually, after replacing the solution for $V_7$, we find an ODE such that the denominator of the functions multiplying the first and zeroth derivatives in $\psi_7$ have three roots, $r_1$, $r_2$, and $r_3$. If the three of them are different from each other, then equation \eqref{psi7} has four singular regular points: $r_1$, $r_2$, $r_3$, and $\infty$. A known function in which the ODE has this property is the Heun function. What we are calling \textit{universal form}, for $\psi_7$ is an ODE in the Sturm-Liouville standard form.\\

We can summarize the results of this section when $\gamma=0$:
\begin{align}\label{gama=0}
\psi_n(\sigma,\eta)&=\left[
f_{\text{AdS}}^4f_\mathcal{I}^{(d_\mathcal{I}-2)/2}
\prod_if_{\mathcal{M}_i}^{d_i/2}\right]^{l/2}.
\end{align}
In the geometry \eqref{a4} we have two spheres and, therefore, two $l$'s, but $\gamma=0$ implies $l_1=l_2\equiv l$, that is why we have only one $l$ in the equation above. 
The operator dimension $\Delta$ when $\gamma=0$ gives
\begin{align}
    \Delta(d=4)=3+2l,\qquad \Delta(d=5)=4+\frac{5l}{2}\label{ld45}\\
    \Delta(d=6)=5+3l,\qquad \Delta(d=7)=6+4l,\label{l67}
\end{align}
and for the solutions with $l\rightarrow -l-1$
\begin{align}
    \Delta(d=4)=2(l+1),\qquad \Delta(d=5)=\frac{5(l+1)}{2}\label{-ld45}\\
    \Delta(d=6)=3(l+1),\qquad \Delta(d=7)=4(l+1).\label{-ld67}
\end{align}
Note that $\Delta(d=5)$ in \eqref{ld45} corresponds to the solution \eqref{n-even} which is for $l$ even, while the one in \eqref{-ld45} corresponds to \eqref{n-odd}, which is for $l$ odd. Therefore, all the $\Delta$ above are positive integers.

\paragraph{AdS$_2$}

An additional solution was obtained after the first version of this article was published on Arxiv. It corresponds to the minimal solution of the class found in \cite{Rigatos:2022ktp}. All the calculations are done in \cite{Rigatos:2022ktp}, but we want to mention that $p(\eta)^\alpha$ is a solution for the $\psi_{l,m,n,p}(\eta)$ function in \eqref{Psi_exp}. The $h$ functions are linear in $\eta$
\begin{align}
    \hat{h}_4(\eta)=a_4\eta+b_4,\qquad
    h_8(\eta)=a_8\eta+b_8.
\end{align}
Differently from the previous cases, here we will not use only the ODE for $V_2$, but its solution:
\begin{equation}
    V_2(\eta)=u(\eta)=\frac{b_0}{2\pi}\eta.
\end{equation}
We then find that $p(\eta)^\alpha$ is solution of \eqref{psi2} with $m^2=l(l+1)$, and therefore $\Delta=l+1$.

\section{Elliptic operators, spectrum and normalizability}\label{norm}

In section \ref{UF and UMS} we found that a specific combination of the warp factors is always a solution in all the cases treated here. However, we did not mention if such solutions are normalizable or not. Let us start with the last one.

\paragraph{AdS$_7$}
Because the warp factors of the solution on AdS$_7$ depend on only one variable, the equation for $\psi_7$ is an ODE, not a PDE. When we write this equation in what we call universal form, we run into an irregular Sturm-Liouville eigenvalue problem. Therefore, a solution of \eqref{psi7} will said to be normalizable if
\begin{align}\label{inner7}
    \int_{\eta_i}^{\eta_{i+1}}w(\eta)|\psi|^2d\eta=
     \int_{\eta_i}^{\eta_{i+1}} f_\text{AdS}^4f_\mathcal{I}^{1/2}f_S|\psi|^2d\eta
      <\infty.
\end{align}
The solution for $V_7$ is straightforward, a third-order polynomial in $\eta$. When we substitute such polynomials in the equation for $\psi_7$, the resulting ODE has at most four singular regular points, the roots of $V_7$ and infinity. If the three roots are different from each other, we have the known Heun differential equation. Let us write $V_7$ as
\begin{align}
    V_7(\eta)=-\frac{162\pi^3F_0}{6}(\eta-a_1)(\eta-a_3)(\eta-a_3),
\end{align}
perform the change of variables
\begin{align}
    \eta=(a_2-a_1)x+a_1,
\end{align}
and rewrite $\psi_7$ as
\begin{align}\label{psi7 H}
    \psi_7=x^{\theta_0}(x-1)^{\theta_1}
    (x-r)^{\theta_r}H(x),\qquad r= \frac{a_1-a_3}{a_1-a_2}.
\end{align}
Then, with $\theta_0=\theta_1=\theta_r=l$, $H(x)$ obeys
\begin{align}
    H''(x)+
    \left[\frac{2(l+1)}{x}+\frac{2(l+1)}{x-1}+
    \frac{2(l+1)}{x-r}
    \right]
    H'(x)+
    \frac{(3x-1-r)(24l+16l^2-m^2)}{x(x-1)(x-r)}H(x)=0
\end{align}
and is solved by
\begin{align}
    &H(x)=c_1Hl(r,q_1;\alpha_1,\beta_1,\gamma,\delta;x)+
    c_2x^{-1-2l}Hl(r,q_2;\alpha_2,\beta_2,\gamma,\delta;x),
    \qquad\text{with}\label{heun}\\
    &q_1=\frac{r+1}{4}(24l+16l^2-m^2),
    \,\,\,\,\,\qquad\qquad\qquad\quad
    q_2=-\frac{r+1}{4}(8+m^2),\\
    &\alpha_1=\frac{1}{2}(5+6l-\sqrt{25-12l(l+1)+3m^2}),\qquad
    \alpha_2=-\alpha_1+2\delta,\\
     &\beta_1=\frac{1}{2}(5+6l+\sqrt{25-12l(l+1)+3m^2})\,\,\,\qquad
      \beta_2=-\beta_1+2\delta,\\
      &\gamma_1=\delta,\qquad\gamma_2=-2l,\qquad
      \delta=2(l+1),
\end{align}
where $Hl(r,q;\alpha,\beta,\gamma,\delta;x)$ is the Heun function.
Remember that the right-hand side of \eqref{V7} is a piecewise constant function. Therefore, \eqref{psi7 H} is a piecewise solution and is normalizable in each interval with the weight function $w(\eta)$. To \eqref{inner7} make sense as an inner product, $w(\eta)$ must be a positive definite function in the range of integration. Therefore, the roots $a_i$ must be such that $w(\eta)\geq 0$ in $[0,\eta_0]$. The inner product between any two solutions is then
\begin{align}
    \left\langle
    \psi_{l_1}|\psi_{l_2}
    \right\rangle =\int_0^{\eta_0} -A^2(\eta-a_1)(\eta-a_2)(\eta-a_3)
    (3\eta-a_1-a_2-a_3) \psi_{l_1}(\eta)\Bar{\psi}_{l_2}(\eta)d\eta.
\end{align}

By requiring analyticity we have to get rid of the second solution, that is $c_2=0$.

When $r=q=0$, $Hl(r,q;\alpha,\beta,\gamma,\delta;x)$ reduces to the hypergeometric function ${}_2F_1(\alpha,\beta;\alpha+\beta-\delta+1,x)$, which correspond to the case when instead of three, $V_7$ has two different roots\footnote{See \cite{Passias:2016fkm}.}. One can easily tell that the UMS for $\psi_7$ we found two sections ago is neither a Heun nor a hypergeometric function, but simply a polynomial function:
\begin{align}
    \psi_{7,\text{UMS}}=\left(\frac{f_{\text{AdS}^4f_S}}{\sqrt{f_\mathcal{I}}}
    \right)^\frac{l}{2}\propto V_7(\eta)^l.
\end{align}
However, when we substitute $m^2$ by \eqref{m7} in \eqref{psi7 H} with $H$ given by \eqref{heun}, the result is the desired $3l$-th order polynomial function.\\

Analogously, the elliptic equations can be written as an eigenvalue problem for an elliptic operator where $m^2$ is the eigenvalue, the weight function is $w$, and then the norm of a solution is given by
\begin{align}
    \int_0^\infty \int_0^\pi\psi\psi^*w(\sigma,\eta)d\eta\,d\sigma.
\end{align}
The normalizability will depend on the solution for $V(\sigma,\eta)$ and on the conditions imposed on it. The potentials $V_4$ and $V_6$ obey the same PDE, which has the following solution
\begin{align}
    V(\sigma,\eta)=\frac{1}{\sigma}\left(
    c_1e^{-\sqrt{c}\sigma}+c_2
    \frac{e^{\sqrt{c}\sigma}}{2\sqrt{c}}
    \right)\left(
    d_1\cos{(\sqrt{c}\,\eta})+d_2\sin{(\sqrt{c}\,\eta)}
    \right),
\end{align}
where $c_1,\,c_2,\,d_1$, and $d_2$ are constants to be fixed by the conditions on $V$, and $c$ is the separation constant. The $w$ and $p$ functions for each cases are
\begin{align}
    w_4&=f_\text{AdS}^4f_\mathcal{I}f_{S_1}f_{S_2},\qquad
    p_4=f_\text{AdS}^4f_{S_1}f_{S_2},\\
    w_6&=f_\text{AdS}^4f_\mathcal{I}f_{S},\qquad\quad\,\,\,\,\,
    p_6=f_\text{AdS}^4f_{S},
\end{align}
and the expressions of the functions above in terms of $V$ are in the Appendix \ref{f functions}, equations \eqref{fads4} and \eqref{fads6}. Let us first analyse the solutions for $\psi_4$. 

\paragraph{AdS$_4$}
The UMS \eqref{g1} writes
\begin{align}\label{psi4-1b}
    \psi_4(\sigma,\eta)&=p_4^{\frac{l^+}{4}}\sigma^{-\frac{l^-}{2}}\nonumber\\
    &=
    \left(  \frac{\pi}{\sqrt{2}}  \right)^{l^+}
    \sigma^{-\frac{l^-}{2}}\left[
    \sigma^2e^{-2\sqrt{c}\,\sigma}
    \left(2\sqrt{c}\,c_1+c_2e^{2\sqrt{c}\,\sigma}\right)^2
    \left(d_2\cos (\sqrt{c}\,\eta)-d_1\sin (\sqrt{c}\,\eta)^2\right)
    \right]^\frac{l^+}{4} ,
\end{align}
and its eigenvalue is $m^2=l^+(l^++3)$.
If we assume $c_1,\,c_2,\,d_1,\, d_2 \in \mathbb{R}$, and $c\geq 0$, then when $\sigma\rightarrow\infty$ the product $\psi_4^2w_4$ diverges if $c_2\neq 0$. That is, if we impose $\psi_4^2w_4<\infty$ when $\sigma\rightarrow\infty$, then $c_2=0$. The eigenvalue depends on $l^+$, therefore the eigenfunctions $\psi_4^{(l_1,l_2)}$ with $(l_1,l_2)=(1,1)$ and $(l_1,l_2)=(0,2)$ have the same eigenvalue. 
After integrating, we find
\begin{align}\label{inner-4}
\left\langle \psi_4^{(l_1,l_2)} \middle|\psi_4^{(\Tilde{l}_1,\Tilde{l}_2)}
\right\rangle=
    \int_0^\pi\int_0^\infty\psi_4^{(l_1,l_2)}\,
   \overline{\psi_4^{(\Tilde{l}_1,\Tilde{l}_2)}}\, w_4\,d\sigma\, d\eta\,=\,
   \delta_{l^+,\Tilde{l}^+}k_{l^+,\Tilde{l}^+}^2,
   \qquad
    \text{if}\qquad
   c=n^2,\quad n\in \mathbb{Z},
\end{align}
and $k_{l^+,\Tilde{l}^+}\in \mathbb{R}$. When $c_2=0$ and $c=n^2$ we have a set of orthogonal functions with positive norm. Therefore, the solutions
\begin{align}\label{psi4-1c}
    \psi_4^{(l^+,n)}(\sigma,\eta)&=p_4^{\frac{l^+}{4}}\sigma^{-\frac{l^-}{2}}\nonumber\\
    &=
    \left(  \frac{\pi}{\sqrt{2}}  \right)^{l^+}
    \sigma^{-\frac{l^-}{2}}\left[4n^2c_1^2
    \sigma^2e^{-2\sqrt{n^2}\,\sigma}
    \left(d_2\cos (\sqrt{n^2}\,\eta)-d_1\sin (\sqrt{n^2}\,\eta)^2\right)
    \right]^\frac{l^+}{4} ,\qquad n\in \mathbb{Z}
\end{align}
are normalizable, and their eigenvalues have degeneracy $l^++1$ and $(l^+-1)/2+1$ when $l^+$ is either even or odd, respectively.

The imposition $c_2=0$ is compatible with either Dirichlet, Neumann or mixed boundary conditions on $\sigma$, since 
\begin{align}
    \psi_4^{(l^+,n)}(0,\eta)=0,\quad
    \psi_4^{(l^+,n)}(\sigma\rightarrow\infty,\eta)=0,\quad
   \left[\partial_\sigma \psi_4^{(l^+,n)}(\sigma,\eta)\right]\Big|_{\sigma=0}=0,\quad
    \left[\partial_\sigma\psi_4^{(l^+,n)}(\sigma,\eta)\right]\Big|_{\sigma\rightarrow\infty}=0.    
\end{align}
While the condition $c=n^2$, $n\in\mathbb{Z}$, is compatible with
\begin{align}
     \psi_4^{(l^+,n)}(\sigma,\pi)=\pm\psi_4^{(l^+,n)}(\sigma,0).
\end{align}
Note that, although $\psi_4^{(l^+,n)}(\sigma,\eta)$ converges on the interval $\sigma\in [0,\infty)$, the same is not true for $V_4(\sigma,\eta)|_{c_2=0}$.\\

\paragraph{AdS$_6$}
The PDE for $V_6$ is the same as for $V_4$; therefore, they have the same solution. When $c_2=0$ and $d_1=0$, $\psi_6$ writes
\begin{align}\label{psi 6 solution}
    \psi_6^{(\Delta_1)}(\sigma,\eta)=A_l\left(
    \frac{(2\sqrt{c}+2\,c\,\sigma)^2\sin^2(\sqrt{c}\,\eta)}{c}
    \right)^{\frac{l}{2}},\quad
    \psi_6^{(\Delta_2)}(\sigma,\eta)=A_l\left(
    \frac{(2\sqrt{c}+2\,c\,\sigma)^2\sin^2(\sqrt{c}\,\eta)}{c}
    \right)^{\frac{l}{2}}\sigma,
\end{align}
and the inner product gives
\begin{align}
\left\langle \psi_6^{(l_1)} \middle|\psi_6^{(l_2)}
\right\rangle=
    \int_0^P\int_0^\infty\psi_6^{(l_1)}\,
   \overline{\psi_6^{(l_2)}}\, w_6\,d\sigma\, d\eta\,=\,
   \delta_{l_1\,\text{mod}\,2,\,l_2\,\text{mod}\,2}C_{l_1,l_2}^2,
   \quad
    \text{with}\quad
   c=\left(\frac{2\pi n}{P}\right)^2,\quad n\in\mathbb{Z},
\end{align}
and the constant $C_{l_1,\,l_2}$ is not the same for the first and second family of solutions in \eqref{psi 6 solution}. 
The imposition $c_2=0$ makes the integration on $\sigma$ convergent. One can not choose either $c_1$ or $d_2$ equal to zero because $w_6(\sigma,\eta)\propto c_1^2d_2^2$.\\

\paragraph{AdS$_5$}
The AdS$_5$ case is a bit more tricky because the solution for the $\sigma$ part of $V_5$ is Bessel functions:
\begin{align}
    V_5(\sigma,\eta)=v_\sigma(\sigma)v_\eta(\eta)=
    \left[
    a_1K_0(\sqrt{c}\,\sigma)+a_1K_0(-\sqrt{c}\,\sigma)\right]\left[
    b_1\cos(\sqrt{c}\,\eta)+b_2\sin(\sqrt{c}\,\eta)
    \right],
\end{align}
where $c$ is, again, the separation constant. It is not tricky in the sense that it is harder to find a solution, we already have it. We may not know the complete solution, but we found that $p^\alpha\sigma^\gamma$ is a solution of the $\psi_5$ equation, and we know $\alpha$ and $\gamma$. Therefore, even if the $V_5$ solution were a Painlevé V times a biconfluent Heun function, we would have the solution for $\psi_5$ anyway. However, the inner product is an integral and such calculation is typically intricate.

Ignoring the symmetry $l\rightarrow-l-1$, we found two solutions for $\psi_5$, and the second one became two. When $n\geq 2(l+1)$ the operator dimension is $\Delta_1^-=n-2l$, which is not in Table 1 and the $\psi_5$ associated with it is not normalizable. On the other hand, when $n\leq 2(l+1)$, $\Delta_1^-=2l-n+4$ is in Table 1, but it is not normalizable either (unless $l=n=0$).

The $\psi_5$ solution with $m^2=m^2_+$ is normalizable and $\Delta_1^+$ is in Table 1. Namely,
\begin{align}\label{psi-5 solution}
    \psi_5=A_{l,n}\left[K_1(\sqrt{c}\,\sigma)\right]^l
    \left[
    b_1\cos(\sqrt{c}\,\eta)+b_2\sin(\sqrt{c}\,\eta)
    \right]^l,
\end{align}
where we already eliminated the solution with the negative argument in the Bessel function because it diverges when $\sigma\rightarrow\infty$. The functions \eqref{psi-5 solution} form an orthogonal set with $c$ given by
\begin{align}
    c=\left(
    \frac{2\pi s}{P} \right)^2,\qquad s\in\mathbb{Z}.
\end{align}
As in \eqref{psi 6 solution}, the solutions \eqref{psi-5 solution} with even $l$ are orthogonal to the answers with odd $l$.


\section{3d-Laplacian}\label{3d-laplacian}

In a certain way, our goal in this paper is to summarize things. We wrote all the spin-2 perturbation parts depending on the $\mathcal{I}$-coordinates in what we called ``universal form", equation \eqref{UF}, and we wrote the simple solutions that depend only on the warp factors, the UMS, for all of them simply with equation \eqref{all-to} and \eqref{ums}. Walking in this ``summarizing things" line, could we resume the final equations found in section \ref{s2-equations}, namely \eqref{psi4}, \eqref{psi5}, \eqref{psi6}, and \eqref{psi7}, as a higher dimensional weighted Laplacian? In other words, we want to know if all of them can come from the following equation:
\begin{equation}\label{L}
    \frac{1}{\sqrt{g}}\partial_\mu(\sqrt{g}g^{\mu\nu}
    \partial_\nu \zeta)=-\lambda^2\omega\zeta,
\end{equation}
where $\omega$ and $\zeta$ are functions of the $\mathcal{I}$ coordinates, while $\lambda$ is a constant.

Let us try a 3-dimensional metric\footnote{One can try a two-dimensional metric, it is simple to show that it does not work.}:
\begin{align}\label{3d metric}
    g_{\mu\nu}=\left(
    \begin{array}{ccc}
f_{11}(\sigma,\eta,r) & f_{12}(\sigma,\eta,r) & f_{13}(\sigma,\eta,r) \\ 
f_{21}(\sigma,\eta,r) & f_{22}(\sigma,\eta,r) & f_{23}(\sigma,\eta,r) \\ 
f_{31}(\sigma,\eta,r) & f_{32}(\sigma,\eta,r) & f_{33}(\sigma,\eta,r)
\end{array} 
\right).
\end{align}
When we substitute it in \eqref{L}, the first condition that emerges is that all off-diagonal terms must vanish. The second condition is and $f_{11}= f_{22}$ . So we will write $g_{\mu\nu}$ as
\begin{align}
    g_{\mu\nu}=\left(
    \begin{array}{ccc}
F_{1}(\sigma,\eta) & 0 & 0 \\ 
0 & F_{1}(\sigma,\eta) & 0 \\ 
0 & 0 & F_{3}(\sigma,\eta,r)
\end{array} 
\right).
\end{align}
Then, equation \eqref{L} becomes
\begin{align}\label{L3}
    \partial_\sigma^2\zeta+\partial_\eta^2\zeta+
    \frac{F_1}{F_3}\partial_r^2\zeta+
    \frac{1}{2F_3}\left(\partial_\sigma F_3 \partial_\sigma\zeta
    + \partial_\eta F_3 \partial_\eta\zeta\right)
    -\frac{F_1\partial_rF_3}{3F_3^2}\partial_r\zeta    +
    \lambda^2F_1\omega\zeta=0.
\end{align}
We will write $F_3$ as $F_3(\sigma,\eta,r)=f_3(\sigma,\eta)f_r(r)$.
Separating variables, $\zeta=\rho(r)P(\sigma,\eta)$, we obtain
\begin{align}\label{PnotSL}
    \frac{1}{F_1\omega}(\partial_\sigma^2+\partial_\eta^2)P+
    \frac{1}{2F_1f_3\omega}\left(
    \partial_\sigma f_3\partial_\sigma+
    \partial_\eta f_3\partial_\eta
    \right)P+\left(
    \frac{a}{f_3\omega}+\lambda^2\right)P=0,
\end{align}
where the separation constant $a$ solves
\begin{align}
    a=\frac{1}{f_r}\frac{\rho''}{\rho}-\frac{1}{2}\frac{f_r'}{f_r^2}
    \frac{\rho'}{\rho},
\end{align}
where the prime means $r$ derivative. Multiplying \eqref{PnotSL} by $F_1\sqrt{f_3}\,\omega$, the equation dresses the form we like:
\begin{align}\label{P}
    \partial_\sigma(p\partial_\sigma P)+
    \partial_\eta(p\partial_\eta P)+ qP=-\lambda^2wP,
\end{align}
where
\begin{align}\label{pqw3d}
    p(\sigma,\eta)=\sqrt{f_3},\qquad
    q(\sigma,\eta)=\frac{a F_1}{\sqrt{f_3}},\qquad
    w(\sigma,\eta)=F_1\sqrt{f_3}\omega.
\end{align}
Comparing \eqref{pqw3d} and \eqref{all-to}, we find
\begin{align}
    f_3&=f_{\text{AdS}_n}^8f_\mathcal{I}^{d_\mathcal{I}-2}
    \prod_i f_{\mathcal{M}_i}^{d_i},\qquad
    F_1=\frac{1}{a}f_{\text{AdS}_n}^8
    f_\mathcal{I}^{d_\mathcal{I}-1}
    \sum_i\frac{\kappa_i}{f_{\mathcal{M}_i}},\\
    \omega&=a\left(
    f_{\text{AdS}_n}^8f_\mathcal{I}^{d_\mathcal{I}-2}
    \prod_i f_{\mathcal{M}_i}^{d_i}
     \sum_j\frac{\kappa_j}{f_{\mathcal{M}_j}}
    \right)^{-1},\qquad \lambda^2=m^2.
\end{align}

\section{Conclusion}\label{conclusion}

Let us close this paper with a brief overview of the main results.

We studied the spin-2 perturbation of four families of 10-dimensional supergravity solutions: two of type IIB on AdS$_4$ and AdS$_6$, with $\mathcal{N}=4$ and $\mathcal{N}=1$, respectively, and two of type IIA on AdS$_5$ and AdS$_7$ with $\mathcal{N}=2$ and $\mathcal{N}=(1,0)$.  All these backgrounds preserve 8 supercharges. The spin-2 perturbation is a natural starting point to study the spectrum because for this perturbation the linearized equations decouple. After some manipulations, we end up with scalar equations depending only on the coordinates of the internal space.
We showed that all the equations on $\mathcal{I}$-coordinates have the same form in terms of the warp factors, which we called \textit{universal form}, equation \eqref{UF} with \eqref{all-to}. After that, we found that $p^\alpha\sigma^\gamma$ (or just $p^\alpha$ in the AdS$_7$ case), with $p(\sigma,\eta)$ as given in \eqref{all-to}, is always a solution for specific values of $\alpha$, $\gamma$, and $m^2$. In other words, we found that for all four backgrounds, there is a combination of the warp factors that is a solution of our eigenvalue problem and we know the corresponding spectrum. Such solutions are called \textit{universal minimal solutions}.

Except for AdS$_7$, we did not find the most general solution of \eqref{UF}. When we compare with the results in \cite{Chen:2019ydk,Gutperle:2018wuk,Passias:2016fkm}, our UMS correspond to the solutions with minimal allowed value for $m^2$. Since we did not use any specific form for the function $V$ until section \ref{norm}, all the results until there are valid for any $V$.

Although all the simple combinations of the warp factors which we found are solutions, they could be non-physical, in the sense that they could turn out to be not real or not normalizable functions. Therefore, in section \ref{norm} we studied the normalizability of the UMS. At this point, we had to use specific expressions for the functions $V$, and from that, we found conditions which make the UMS normalizable.  By ``normalizable'' we mean that the norm of the solutions, seen as eigenfunctions of an elliptic operator (when we are talking about solutions on AdS$_{4,5,6}$), or a Sturm-Liouville (when on AdS$_7$) eigenvalue problem, is finite.
By demanding analyticity and normalizability we obtained conditions for the parameters. The first requirement was done appeared in the $\sigma$ variable of the function $\psi$, and the second in the $\eta$ part. For $\psi_7$ the problem was handled as a Sturm-Liouville eigenvalue problem.  The solutions for $\psi$ are labelled with some parameters, which means that in each case we have a set of solutions. Then, we found the values of the separation constant $c$ for which the set of solutions is orthogonal. 
We also found that for all the solutions the operator dimension obeyed unitarity condition, as showed in \cite{Cordova:2016emh}, are normalizable. In appendix \ref{algebra}, we identified the multiplets containing the stress tensor.  

 The way the spectra were found in this work is somewhat nonstandard. In particular, we did not explicitly impose boundary conditions to fix the solution for $\psi$ and, consequently, find the spectrum. After imposing analyticity, normalizability, and orthogonality of the solutions in section \ref{norm}, we found an expression for $\psi_4$, which still has two free parameters. After equation \eqref{psi4-1a}, we discussed boundary conditions compatible with our $\psi_4$. We expect that the explicit boundary conditions are chosen by the physics of the specific problem. In particular, the discussed geometries are generated by D-branes, which themselves must impose appropriate boundary conditions. The details of such boundary conditions will be discussed elsewhere.

Using either the standard or the non-standard way of finding spectra, it is important to know properties and theorems about the corresponding eigenvalue problems. For this reason, future work is to understand how theorems about the spectrum of elliptic operators in unbounded domains, as in \cite{Lewis-elliptic} for example, can help us to understand the boundary conditions and their implications on the physics. Another future work is to investigate the dual theory, establishing exact correspondence between the supergravity backgrounds and superconformal field theories.

\section*{Acknowledgements}

I want to thank Carlos Núñez for having suggested this project and for guiding me. Thiago Fleury, Dmitry Melnikov, Daniel Thompson, and Alessandro Tomasiello for all the discussions and suggestions. Swansea University, where I obtained most of the results presented here, has received me as a student visitor. This work was supported by CAPES, the Serrapilheira Institute (grant number Serra – R-2012-38185), and the Simons Foundation 1023171-RC.

\appendix

\section{f functions - the translation}\label{f functions}

All the metrics in the present paper are 10-dimensional and have the following form

\begin{align}
    ds^2_{10,\, Einstein}=f_{\text{AdS}}\left[
ds^2(\text{AdS}_n)+f_{S_1}ds^2(S_1^2)+f_{S_2}ds^2(S_2^2)+f_\beta d\beta^2+f_\mathcal{I}ds^2(\mathcal{I})
    \right]
\end{align}

We can see in the table below the expressions for the warp factors and the respective references we are using.

\begin{center}
\begin{tabular}{|c|c|c|c|c|c|c|}
\hline 
 $AdS_n$ & $f_\text{AdS}$ & $f_{S_1}$ & $f_{S_2}$ & $f_\beta$ & $f_\mathcal{I}$ & Ref. \\ 
\hline 
4 & $f_1f_5^{1/4}$ & $f_2$ & $f_3$ & 0 & $f_4$ & \cite{Ricardo} \\ 
\hline 
5 & $e^{-\frac{\Phi}{2}}f_0$ & $f_1$ & 0 & $f_3$ & $f_2$ & \cite{Itsios:2019yzp} \\ 
\hline 
6 & $f_6^{1/4}f_1$ & $f_2$ & 0 & 0 & $f_3$ & \cite{Legramandi:2021uds} \\ 
\hline 
7 & $f_1$ & $f_3/f_\text{AdS}$ & 0 & 0 & $f_2/f_\text{AdS}$ & \cite{Nunez:2018ags} \\ 
\hline 
\end{tabular} 
\end{center}

AdS$_4$:
\begin{align}\label{fads4}
    f_1&=\frac{\pi}{2}\sqrt{\frac{\sigma^3\partial_{\sigma\eta}^2V}{\partial_\sigma(\sigma\partial_\eta V)}},\qquad
    f_2=-\frac{\partial_\eta V\partial_\sigma(\sigma\partial_\eta V)}{\sigma\Lambda},\qquad
    f_3=\frac{\partial_\sigma(\sigma\partial_\eta V)}{\sigma\partial_{\eta\sigma}^2V},\qquad
    f_4=-\frac{\partial_\sigma(\sigma\partial_\eta V)}{\sigma^2\partial_\eta V}\nonumber \\
    f_5&=-16\frac{\Lambda\partial_\eta V}{\partial^2_{\eta\sigma}V},\qquad
    \Lambda=\partial_\eta V\partial^2_{\eta\sigma}V+
    \sigma [(\partial_{\eta\sigma}^2V)^2+(\partial_\eta^2V)^2].
\end{align}
See \cite{Ricardo} to see the other $f$ functions and the $B$, $C$, $F$, and $H$  fields expressions.\\

AdS$_5$:
\begin{align}\label{fads5}
    &f_0=\sigma\sqrt{\frac{\partial_\sigma(\sigma\partial_\sigma V)-
    2\partial_\sigma V}{\partial_\sigma(\sigma\partial_\sigma V)}},\qquad
    f_1=-\frac{\partial_\sigma(\sigma\partial_\sigma V)
    \partial_\sigma V}{2\Delta},\qquad
    f_2=-\frac{\partial_\sigma(\sigma\partial_\sigma V)}
    {2\sigma^2\partial_\sigma V}\nonumber \\
    &f_3= \frac{\partial_\sigma(\sigma\partial_\sigma V)}
    {\partial_\sigma(\sigma\partial_\sigma V)-2\partial_\sigma V},\qquad
    e^{2\Phi}=2^7\frac{(\sigma\partial_\sigma(\sigma\partial_\sigma V)-
    2\sigma\partial_\sigma V)^{3/2}}{\sqrt{\sigma\partial_\sigma(\sigma\partial_\sigma V)}
    \partial_\sigma V \Delta},\nonumber \\
    &\Delta=(\partial_\sigma(\sigma\partial_\sigma V)-
    2\partial_\sigma V)\partial_\sigma(\sigma\partial_\sigma V)+
    (\sigma\partial_\sigma\partial_\eta V)^2.
\end{align}
See \cite{Itsios:2019yzp} to see the NS and RR field expressions.\\

AdS$_6$:

\begin{align}\label{fads6}
    &f_1=\frac{2}{3}\sqrt{\sigma^2+\frac{3\sigma\partial_\sigma V}
    {\partial_\eta^2V}},\qquad
    f_2=\frac{\partial_\sigma V\partial_\eta^2V}{3\Lambda},\qquad
    f_3= \frac{\partial_\eta^2V}{3\sigma\partial_\eta V},\nonumber\\
    &\Lambda = \sigma(\partial_\eta \partial_\sigma V)^2+
    (\partial_\sigma V-\sigma\partial_\sigma^2V)\partial_\eta^2V, \qquad
    f_6= 18^2\frac{3\sigma^2\partial_\sigma V\partial_\eta^2V}{(3\partial_\sigma V+\sigma\partial_\eta^2 V)^2}\Lambda.
\end{align}\\

AdS$_7$:
\begin{align}\label{fads7}
    &f_1=8\sqrt{2}\pi \sqrt{-\frac{\alpha}{\alpha''}},\qquad
    f_2 = \sqrt{2}\pi \sqrt{-\frac{\alpha''}{\alpha}},\qquad
    f_3 = \sqrt{2}\pi \sqrt{-\frac{\alpha''}{\alpha}}\left(
    \frac{\alpha^2}{\alpha'^{2}-2\alpha\alpha''}
    \right),\nonumber\\
    &f_6=2^{\frac{5}{4}}\pi^\frac{5}{2}3^4
    \frac{(-\alpha\alpha'')^\frac{3}{4}}{\alpha'^2-2\alpha\alpha''},\qquad 
    \text{where} \qquad \alpha = V_7(\eta).
\end{align}


\section{A bit of the algebra part}\label{algebra}
The goal of this appendix is to make a bridge between our results and \cite{Cordova:2016emh}. Therefore, we will use their conventions. 
In \cite{Cordova:2016emh}, the authors made an extensive study of multiplets of superconformal algebras in\footnote{In this appendix we will denote by $d$ the dimension of the superconformal algebra. That is the AdS dimension minus one. } $3\geq d\geq 6$. 
Given the properties of our backgrounds, such a bridge is to recognize which operator dimension evaluated here are in Tables 6, 13, 14, 22, and 23 in their paper.

The $\mathcal{N}=4$ superconformal algebra in $d=3$ is the $\mathfrak{osp}(4|4)$. Since in the present paper we did spin-2 perturbation, the $\mathfrak{su}(2)$ Dynkin label $j$ is then $j=4$, while the charges $R$ and $R'$ are $2l_1$ and $2l_2$. Analogous, for $\mathcal{N}=2$ in $d=4$, $\mathcal{N}=1$ in $d=5$, and for $\mathcal{N}=(1,0)$ in $d=6$, the spin-2 perturbation correspond to an state with Dynkin labels $[j,\Bar{j}]=[2,2]$, $[j_1,j_2]=[0,2]$, and $[j_1,j_2,j_3]=[0,2,0]$, respectively; the angular charge is $R=2l$, and the u(1) charge is $r=\pm n$. See Tables \ref{t-rep} and \ref{t-delta}.

As expected, the operator dimensions matching the inequalities in \cite{Cordova:2016emh} are associated with normalizable solutions. We know that the stress tensor corresponds to a state with $\Delta=d$ and no charge. Besides that, it must be the top component of a short multiplet. This is exactly what we found and we draw the stress multiplet for each case. See below.

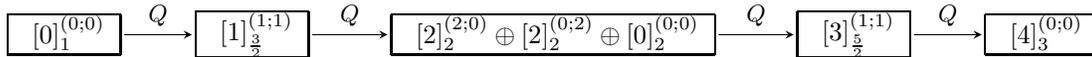
\begin{figure}[!htbp]
\begin{center}
\begin{tikzpicture}[scale=1]
  \node[] at (0,0) { \begin{tabular}{|l|}
\hline 
\rule[-1ex]{0pt}{2.5ex} $[2]_2^{(2;0)}\oplus [2]_2^{(0;2)}\oplus [0]_2^{(0;0)}$ \\ 
\hline 
\end{tabular} };
\node[] at (-4,0) { \begin{tabular}{|l|}
\hline 
\rule[-1ex]{0pt}{2.5ex} $[1]_\frac{3}{2}^{(1;1)}$ \\ 
\hline 
\end{tabular} };
\node[] at (4,0) { \begin{tabular}{|l|}
\hline 
\rule[-1ex]{0pt}{2.5ex} $[3]_\frac{5}{2}^{(1;1)}$ \\ 
\hline 
\end{tabular} };
\node[] at (-6.5,0) { \begin{tabular}{|l|}
\hline 
\rule[-1ex]{0pt}{2.5ex} $[0]_1^{(0;0)}$ \\ 
\hline 
\end{tabular} };
\node[] at (6.5,0) { \begin{tabular}{|l|}
\hline 
\rule[-1ex]{0pt}{2.5ex} $[4]_3^{(0;0)}$ \\ 
\hline 
\end{tabular} };
\draw[-stealth](-5.7,0) -- (-4.8,0);
\draw[-stealth](-3.2,0) -- (-2.2,0);
\draw[-stealth](4.8,0) -- (5.7,0);
\draw[-stealth](2.2,0) -- (3.2,0);
\node[] at (-5.25,0.25) {\small $Q$};
\node[] at (5.25,0.25) {\small $Q$};
\node[] at (-2.7,0.25) {\small $Q$};
\node[] at (2.7,0.25) {\small $Q$};
  \end{tikzpicture}\caption{$A_2[0]_1^{(0;0)}$, $(d=3,\, \mathcal{N}=4)$. See (5.55) in \cite{Cordova:2016emh}. }\label{34}
\end{center}
\end{figure}

All the multiples below are superconformal current multiples. More precisely, they are the stress tensor multiples. Therefore, each of them contains the stress tensor $T_{\mu\nu}$ and its superpartners. $T_{\mu\nu}$ is the top component of a superconformal multiple which also contains $S_{\mu\alpha}$ and $j_\mu^{(R)}$, the $S$-superchargers and the $R$-current, respectively. While $T_{\mu\nu}$ has dimension $\Delta_{T_{\mu\nu}}=d$, for $S_{\mu\alpha}$ and $j_\mu^{(R)}$ we have $\Delta_{S_{\mu\alpha}}=d-1/2$ and $\Delta_{j_\mu^{(R)}}=d-1$. Then, for example, we can then identify that in Figure \ref{34} we have, from the right to the left, $T_{\mu\nu}$,  $S_{\mu\alpha}$, and $j_\mu^{(R)}$.

\begin{center}
\begin{table}[!htbp]
\centering
\begin{tabular}{|c|c|c|c|c|c|c|} 
\hline 
$(d,\,\mathcal{N})$ & Algebra &  Representation  & Charges \\ 
\hline 
 (3, 4)& $\mathfrak{osp}(4|4)$ & $[j]_\Delta^{(R;R')}=[4]_{\Delta_1}^{(2l_1;2l_2)}$ & $R=2l_1,\quad R'=2l_2$  \\ 
\hline 
(4, 2) & $\mathfrak{su}(2,2|2)$ & $[j,\Bar{j}]_\Delta^{(R;r)}=[2,2]_{\Delta_1^{++}}^{(2l;r)}$  & $R=2l,\quad r=-2n$\\ 
(4, 2) & $\mathfrak{su}(2,2|2)$ & $[j,\Bar{j}]_\Delta^{(R;r)}=[2,2]_{\Delta_1^{++}}^{(2l;r)}$  & $R=2l,\quad r=2n$\\ 
(4, 2) & $\mathfrak{su}(2,2|2)$ & $[j,\Bar{j}]_\Delta^{(R;r)}=[2,2]_{\Delta_1^{-}}^{(2l;r)}$  & $R=2l,\quad  r=2n$, $n\leq 2(l+1)$ \\ 
(4, 2) & $\mathfrak{su}(2,2|2)$ & $[j,\Bar{j}]_\Delta^{(R;r)}=[2,2]_{\Delta_1^{-}}^{(2l;r)}$  & $R=2l,\quad  r=-2n$, $n\leq 2(l+1)$ \\ 
\hline
(5, 1) & $\mathfrak{so}(5)$  & $[j_1,j_2]_\Delta^{(R)}=[0,2]_{\Delta_1}^{(2l)}$& $R=2l$ \\ 
(5, 1) & $\mathfrak{so}(5)$  & $[j_1,j_2]_\Delta^{(R)}=[0,2]_{\Delta_2}^{(2l)}$ & $R=2l$\\ 
\hline
(6, $\mathcal{N}=(1,0)$) & $\mathfrak{so}(6)$  & $[j_1,j_2,j_3]_\Delta^{(R)}=[0,2,0]_{\Delta_1}^{(2l)}$  & $R=2l$ \\ 
\hline
\end{tabular}
\caption{The corresponding algebra, representation and respective charges.}
   \label{t-rep}
\end{table}
\end{center}

\begin{center}
\begin{table}[!htbp]
\centering
\begin{tabular}{|c|c|c|c|c|c|c|}
\hline 
$(d,\,\mathcal{N})$ &  SC type & $\Delta$ & Primary\\
\hline 
 (3, 4) & $A_1$  & $\Delta_1=l^++3$ & $[4]_{\Delta_1}^{(2l_1;2l_2)}$\\
\hline 
(4, 2)  & $A_1$& $\Delta_1^{+}=2l+n+4$ & $[2,2]_{\Delta^{++}_1}^{(2l;\, r)}$\\
(4, 2)  & $\Bar{A}_1$ & $\Delta_1^{+}=2l+n+4$ & $[2,2]_{\Delta^{++}_1}^{(2l;\, r)}$ \\
(4, 2)  & $A_1$  & $\Delta_1^{-}=2l-n+4$ & $[2,2]_{\Delta^{-}_1}^{(2l;\, r)}$\\
(4, 2) & $\Bar{A}_1$  & $\Delta_1^{-}=2l-n+4$ & $[2,2]_{\Delta^{-}_1}^{(2l;\, r)}$\\
\hline
(5, 1)  & $ B_1$  & $\Delta_1=3l+5$ & $[0,2]_{\Delta_1}^{(2l)}$ \\
(5, 1) & $A_2$  & $\Delta_2=3l+6$ & $[0,2]_{\Delta_2}^{(2l)}$ \\
\hline
(6, $\mathcal{N}=(1,0)$) & $B_1$ & $\Delta_1=4l+6$ & $[0,2,0]_{\Delta_1}^{(2l)}$ \\
\hline
\end{tabular}
\caption{The shortening condition (SC) type obeyed by the operator dimension, $\Delta$, and the primary for each unitary solution found.}
     \label{t-delta}
\end{table}
\end{center}

\vspace{0.5in}

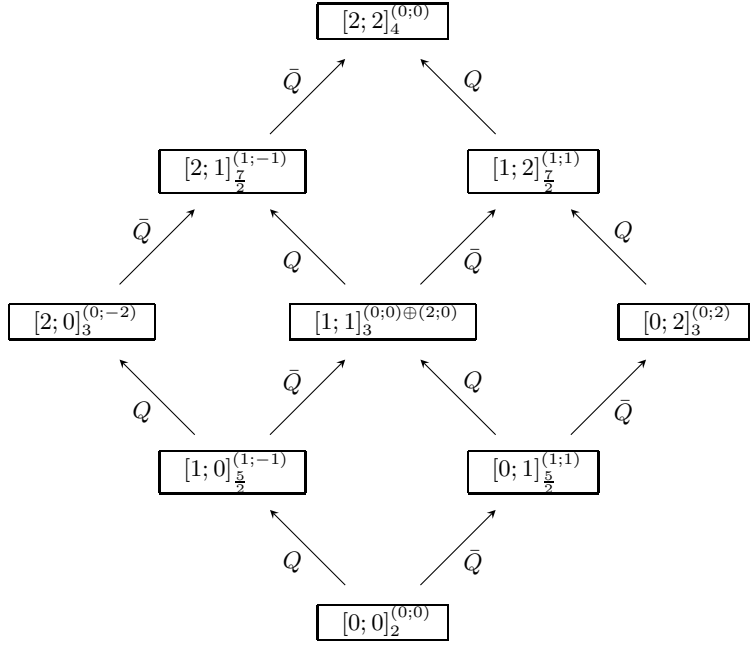
\begin{figure}[!htbp]
\begin{center}
\begin{tikzpicture}[scale=1]
  \node[] at (0,0) {\small \begin{tabular}{|l|}
\hline 
\rule[-1ex]{0pt}{2.5ex} $[1;1]_3^{(0;0)\oplus(2;0)}$ \\ 
\hline 
\end{tabular} };
\node[] at (0,4) {\small \begin{tabular}{|l|}
\hline 
\rule[-1ex]{0pt}{2.5ex} $[2;2]_4^{(0;0)}$ \\ 
\hline 
\end{tabular} };
\node[] at (0,-4) {\small \begin{tabular}{|l|}
\hline 
\rule[-1ex]{0pt}{2.5ex} $[0;0]_2^{(0;0)}$ \\ 
\hline 
\end{tabular} };
\node[] at (-2,2) {\small \begin{tabular}{|l|}
\hline 
\rule[-1ex]{0pt}{2.5ex} $[2;1]_\frac{7}{2}^{(1;-1)}$ \\ 
\hline 
\end{tabular} };
\node[] at (2,2) {\small \begin{tabular}{|l|}
\hline 
\rule[-1ex]{0pt}{2.5ex} $[1;2]_\frac{7}{2}^{(1;1)}$ \\ 
\hline 
\end{tabular} };
\node[] at (-4,0) {\small \begin{tabular}{|l|}
\hline 
\rule[-1ex]{0pt}{2.5ex} $[2;0]_3^{(0;-2)}$ \\ 
\hline 
\end{tabular} };
\node[] at (4,0) {\small \begin{tabular}{|l|}
\hline 
\rule[-1ex]{0pt}{2.5ex} $[0;2]_3^{(0;2)}$ \\ 
\hline 
\end{tabular} };
\node[] at (-2,-2) {\small \begin{tabular}{|l|}
\hline 
\rule[-1ex]{0pt}{2.5ex} $[1;0]_\frac{5}{2}^{(1;-1)}$ \\ 
\hline 
\end{tabular} };
\node[] at (2,-2) {\small \begin{tabular}{|l|}
\hline 
\rule[-1ex]{0pt}{2.5ex} $[0;1]_\frac{5}{2}^{(1;1)}$ \\ 
\hline 
\end{tabular} };
\draw[-stealth](-1.5,2.5) -- (-0.5,3.5);
\draw[-stealth](1.5,2.5) -- (0.5,3.5);
\draw[-stealth](-3.5,0.5) -- (-2.5,1.5);
\draw[-stealth](3.5,0.5) -- (2.5,1.5);
\draw[-stealth](-0.5,0.5) -- (-1.5,1.5);
\draw[-stealth](0.5,0.5) -- (1.5,1.5);
\draw[-stealth](-2.5,-1.5) -- (-3.5,-0.5);
\draw[-stealth](2.5,-1.5) -- (3.5,-0.5);
\draw[-stealth](-1.5,-1.5) -- (-0.5,-0.5);
\draw[-stealth](1.5,-1.5) -- (0.5,-0.5);
\draw[-stealth](-0.5,-3.5) -- (-1.5,-2.5);
\draw[-stealth](0.5,-3.5) -- (1.5,-2.5);
 \node[] at (-1.2,-3.2) {\small $Q$};
 \node[] at (1.2,-3.2) {\small $\Bar{Q}$};
 \node[] at (-1.2,-0.8) {\small $\Bar{Q}$};
 \node[] at (1.2,-0.8) {\small $Q$};
 \node[] at (-1.2,0.8) {\small $Q$};
 \node[] at (1.2,0.8) {\small $\Bar{Q}$};
 \node[] at (-1.2,3.2) {\small $\Bar{Q}$};
 \node[] at (1.2,3.2) {\small $Q$};
  \node[] at (-3.2,1.2) {\small $\Bar{Q}$};
  \node[] at (3.2,1.2) {\small $Q$};
  \node[] at (-3.2,-1.2) {\small $Q$};
 \node[] at (3.2,-1.2) {\small $\Bar{Q}$};
\end{tikzpicture}\caption{$A_2\Bar{A}_2$, $(d=4,\,\mathcal{N}=2)$.  See (5.95) in \cite{Cordova:2016emh}.}
\label{42}
\end{center}
\end{figure}


\begin{figure}[!htbp]
\begin{center}
\begin{tikzpicture}[scale=1]
 \node[] at (0,0) { \begin{tabular}{|l|}
\hline 
\rule[-1ex]{0pt}{2.5ex} $[0;1]_4^{(2)}\oplus [2;0]_4^{(0)}$ \\ 
\hline 
\end{tabular} };
 \node[] at (-3.8,0) { \begin{tabular}{|l|}
\hline 
\rule[-1ex]{0pt}{2.5ex} $[1;0]_\frac{7}{2}^{(1)}$ \\ 
\hline 
\end{tabular} };
\node[] at (-7,0) { \begin{tabular}{|l|}
\hline 
\rule[-1ex]{0pt}{2.5ex} $[0;0]_3^{(0)}$ \\ 
\hline 
\end{tabular} };
 \node[] at (3.8,0) { \begin{tabular}{|l|}
\hline 
\rule[-1ex]{0pt}{2.5ex} $[1;1]_\frac{9}{2}^{(1)}$ \\ 
\hline 
\end{tabular} };
\node[] at (7,0) { \begin{tabular}{|l|}
\hline 
\rule[-1ex]{0pt}{2.5ex} $[0;2]_5^{(0)}$ \\ 
\hline 
\end{tabular} };
\draw[-stealth](-6.1,0) -- (-4.7,0);
\draw[-stealth](4.7,0) -- (6.1,0);
\draw[-stealth](-2.9,0) -- (-1.6,0);
\draw[-stealth](1.6,0) -- (2.9,0);
\node[] at (-5.4,0.25) {\small $Q$};
\node[] at (5.4,0.25) {\small $Q$};
\node[] at (2.25,0.25) {\small $Q$};
\node[] at (-2.25,0.25) {\small $Q$};
\end{tikzpicture}\caption{$B_2[0,0]_3^{(0)}$, $(d=5,\, \mathcal{N}=1)$.  See (5.111) in \cite{Cordova:2016emh}.}
\label{51}
\end{center}
\end{figure}
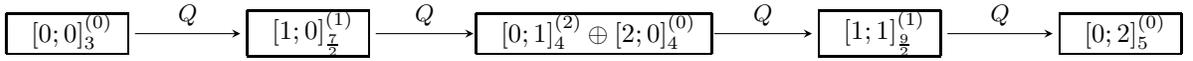

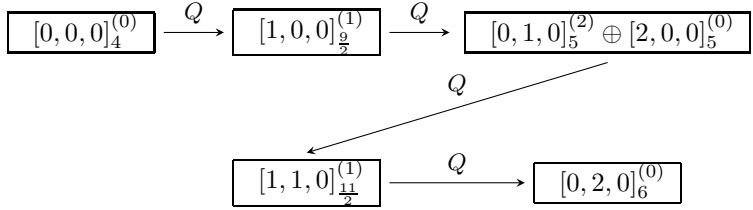
\begin{figure}[!htbp]
\begin{center}
\begin{tikzpicture}[scale=1]
 \node[] at (0,0) { \begin{tabular}{|l|}
\hline 
\rule[-1ex]{0pt}{2.5ex} $[1,0,0]_\frac{9}{2}^{(1)}$ \\ 
\hline 
\end{tabular} };
\node[] at (-3,0) { \begin{tabular}{|l|}
\hline 
\rule[-1ex]{0pt}{2.5ex} $[0,0,0]_4^{(0)}$ \\ 
\hline 
\end{tabular} };
\node[] at (4,0) { \begin{tabular}{|l|}
\hline 
\rule[-1ex]{0pt}{2.5ex} $[0,1,0]_5^{(2)}\oplus [2,0,0]_5^{(0)}$ \\ 
\hline 
\end{tabular} };
\node[] at (0,-2) { \begin{tabular}{|l|}
\hline 
\rule[-1ex]{0pt}{2.5ex} $[1,1,0]_\frac{11}{2}^{(1)}$ \\ 
\hline 
\end{tabular} };
\node[] at (4,-2) { \begin{tabular}{|l|}
\hline 
\rule[-1ex]{0pt}{2.5ex} $[0,2,0]_6^{(0)}$ \\ 
\hline 
\end{tabular} };
\draw[-stealth](-1.9,0) -- (-1.1,0);
\draw[-stealth](1.1,0) -- (1.9,0);
\draw[-stealth](4,-0.4) -- (0,-1.6);
\draw[-stealth](1.1,-2) -- (2.9,-2);
\node[] at (-1.5,0.25) {\small $Q$};
\node[] at (1.5,0.25) {\small $Q$};
\node[] at (2,-0.7) {\small $Q$};
\node[] at (2,-1.75) {\small $Q$};
\end{tikzpicture}\caption{$B_3[0,0,0]_4^{(0)}$, $(d=6,\, \mathcal{N}=(1,0))$.  See (5.128) in \cite{Cordova:2016emh}.}
\label{610}
\end{center}
\end{figure}
\vspace{1in}

\printbibliography

\end{document}